\documentclass[11pt]{article}

\usepackage{amssymb}
\usepackage{amsfonts}
\usepackage{latexsym}

\usepackage{subfigure}
\usepackage[dvips,final]{epsfig}
\usepackage{color}

\newcommand{\qa}{\mbox{\quad\mbox{and}\quad}}

\newcommand{\g}{\mbox{$\mathcal G$}}

\newcommand{\kk}{\mbox{$\mathcal K$}}

\newcommand{\oo}{\mbox{$\mathbb O$}}

\newcommand{\x}{\mbox{$\mathbf x$}}
\newcommand{\y}{\mbox{$\mathbf y$}}
\newcommand{\vv}{\mbox{$\mathbf v$}}

\newcommand{\zz}{\mbox{$\mathbf z$}}

\newcommand{\rt}{\mbox{$\mathbb R$}}

\newcommand{\ttt}{\mbox{$\mathcal T$}}
\newcommand{\f}{\mbox{$\mathcal F$}}

\newcommand{\p}{\mbox{$\mathcal P$}}
\newcommand{\q}{\mbox{$\mathcal Q$}}

\newcommand{\rrr}{\mbox{$\mathcal R$}}

\newcommand{\ww}{\mbox{$\mathbf w$}}

\newcommand{\et}{\mbox{$\mbox{\boldmath $\eta$}$}}
\newcommand{\xxi}{\mbox{$\mbox{\boldmath $\xi$}$}}

\newcommand{\tth}{\mbox{$\mbox{\boldmath $\theta$}$}}

\newtheorem{definition}{Definition}
\newtheorem{remark}{Remark}
\newtheorem{theorem}{Theorem}
\newtheorem{lemma}{Lemma}
\newtheorem{corollary}{Corollary}
\newtheorem{example}{Example}

\date{}

\begin{document}
 \title{ Interpolation Estimator for Infinite Sets of Random Vectors}

 \author{ A. Torokhti  \\
  {\small {\em Centre for Industrial and Applied  Mathematics} }
 \\{\small {\em   University of South  Australia, SA 5095, Australia } }
 \\{\small  anatoli.torokhti@unisa.edu.au} }

\maketitle

\begin{abstract}
We propose an approach to the filtering of  infinite sets  of stochastic signals, { $K_{_Y}$ and $K_{_X}$.
 The known Wiener-type approach cannot be applied to infinite sets of signals.  Even in the case when $K_{_Y}$ and
$K_{_X}$ are finite sets, the computational
work associated with the Wiener approach becomes unreasonably hard. To avoid such difficulties,} a new  theory
is studied.

The problem addressed is as follows.
Given two {infinite} sets of stochastic signals, $K_{_Y}$ and $K_{_X}$, find a single filter $\f: K_{_Y}\rightarrow K_{_X}$
that estimates signals from $K_{_Y}$ with a controlled associated error.

Our approach is based on exploiting a signal interpolation idea. The proposed filter $F$ is represented in the form of a sum
of $p$ terms, $F(\y) =  \sum_{j=1}^{p}T_j R_j Q_j (\y) $. Each term is
derived from three operations presented by matrices, $Q_i$, $R_i$ and $T_i$ with $i=1,\ldots,p$. Each operation is a
special stage of the filtering aimed at facilitating the associated numerical work.
In particular,  $\q_1$, $\ldots,$ $\q_{p}$ are used to transform an observable signal $\y\in K_{_Y}$
to $p$ different signals. Matrices $R_1,\ldots,R_p$ reduce a set of related matrix equations  to $p$ {\em independent}
equations.  Their solution requires
much less computational effort than would be required with the full set of matrix equations. Matrices $T_i,\ldots,T_p$
are determined from {\em interpolation conditions.} We show that the proposed filter is asymptotically optimal.
Moreover, the filter model is determined in terms of pseudo-inverse matrices  and, therefore, it always exists.
\end{abstract}

\section{Introduction}\label{intro}

\subsection{Motivation and basic idea}\label{motiv}

In this paper, we consider extensions of the known approaches to filtering of random signals based on the Wiener idea
\cite{tor100}. Wiener-type filters have a broad spectrum of applications in engineering, physics, biology and many
other research areas.
Some relevant references can be found, for example, in \cite{gao1,hay2,jeo1,red1}.

\subsubsection{Motivation}\label{mot}

One  motivation for the proposed method is as follows. Most of the literature on the subject of Wiener-type
filtering\footnote{Relevant references can be found, for example, in \cite{tor100}.} discusses the properties of an
optimal filter for an {individual} {\sf\em finite} random signal-vector.\footnote{We say that a random signal-vector $\x$
is finite if $\x$ has a finite number of components. } This means that if one wishes to { transform} an {infinite}
 set of observable random signal-vectors $K_{_Y}=\{\y_1, \y_2, \ldots, \y_N, \ldots\}$ { so that it should
be close to an {infinite}  set of reference random signal-vectors $K_{_X}=\{\x_1, \x_2, \ldots, \x_N, \ldots\}$ }
using the standard Wiener-type approach then one is forced to find an {infinite} set of corresponding Wiener
filters $\{\f_1, \f_2, \ldots,\f_N, \ldots\}$: one $\f_i$ of the filter set for each representative $\y_i$ of the
signal set $K_{_Y}$. Therefore, if $K_{_Y}$ and $K_{_X}$ are infinite sets of random signals, such an approach cannot be
applied in practice.\footnote{Here, $K_{_Y}$ and $K_{_X}$ are countable sets. More generally, the sets $K_{_Y}$ and
$K_{_X}$ might be uncountable  when $K_{_Y}$ and $K_{_X}$ depend on a continuous parameter $\alpha$.}
 Moreover, in some situations it is appropriate and possibly necessary to introduce an `identifier', a parameter whose value
identifies or associates a specific filter from the set $\{\f_1, \f_2, \ldots,\f_N, \ldots\}$ with a given signal
from $K_{_Y}$.

An example of infinite sets, $K_{_Y}$ and $K_{_X}$, arises in the situation where signals $\y\in K_{_Y}$ and
$\x\in K_{_X}$ are dependent on a parameter vector $\alpha = (\alpha^{(1)}, \ldots, \alpha^{(q)})^T \in C^q\subseteq
\rt^q$, where $C^q$ is a $q$-dimensional cube, i.e., $\y = \y(\cdot,\alpha)$ and $\x = \x(\cdot,\alpha)$. In particular,
one coordinate, say $\alpha^{(1)}$ of $\alpha$, could be interpreted as time, thus allowing for a continuous stream of
vectors. In this regard, see also Section \ref{choiceqi} below.

{ Note that even in the case when $K_{_Y}$ and $K_{_X}$ are finite sets, $K_{_Y}=\{\y_1, \y_2, \ldots, $ $\y_N\}$
and $K_{_X}=\{\x_1, \x_2, \ldots, \x_N\}$,  and then $K_{_Y}$ and $K_{_X}$ can be represented as finite signals,
the Wiener approach leads to computation of large covariance matrices.  Indeed, if each $\y_i$ has $n$ components
and each $\x_i$ has $m$ components then the Wiener approach leads to computation of a product of an $mN \times
nN$ matrix and an $nN \times nN$ matrix and computation of an $nN \times nN$ pseudo-inverse matrix \cite{tor100}.
This requires $O(2mn^2N^3)$ and $O(22n^3N^3)$ flops, respectively \cite{gol1}. As a result, the computational work associated
with this approach becomes unreasonably hard.}

{ Note that a filtering methodology based on recursive least squares (RLS) algorithms,
similar to Wiener-type filters, also discusses the properties of a filter for an { individual} {finite}  signal-vector,
not for infinite signal sets considered in this paper  (see, for example, \cite{ale1,apo1,hay1,say1}). Additionally, in the derivation of  RLS filters, it is  assumed
 that the correlation matrix is invertible\footnote{Sections 9.2--9.3 in \cite{hay1}.}. In our method, the latter restriction is omitted
 (see  Remark \ref{ps-inv}). It is also known \cite{say1} that the computational cost of RLS filters is of the same order or  higher
 than that of the  Wiener-type filtering techniques.}

\subsubsection{Brief description of the problem}

To avoid such difficulties, we here propose and study an approach that allows one to use a {single} filter
to estimate arbitrary signals from the set $K_{_Y}$. In other words, we solve the following problem. Given two
{infinite} sets of random signals, $K_{_Y}$ and $K_{_X}$, find a single filter $\f: K_{_Y}\rightarrow K_{_X}$
based on an observable signal $\y \in K_{_Y}$ that estimates the signal $\x \in K_{_X}$ with a controlled, associated
error. Note that in our formulation the set $K_{_Y}$ can be finite or infinite. In the latter case, though,
the set must be compact.

\subsubsection{Basic idea}\label{basic}

Our proposed approach to determining the desired filter $\f:K_{_Y}\rightarrow K_{_X}$ is as follows. First, choose a
finite subset $S_{_X}=\{\x_1, \x_2, \ldots, \x_p\}\subset K_{_X}$ and a corresponding subset $S_{_Y}=\{\y_1, \y_2,
\ldots,$ $\y_p\}\subset K_{_Y}$. Then, define $p$ pairs $(\x_1,\y_1),\ldots, (\x_p,\y_p)$ taken from the set $S_{_X}
\times S_{_Y}$. These are used to establish $p$ interpolation conditions. Finally, determine the filter $\f$ that
satisfies the interpolation conditions. The filter $\f$  is called the interpolation filter.

\subsubsection{Contribution}\label{contr}

In Section \ref{err}, we show that the interpolation  filter $\f$ is asymptotically optimal
in the sense that, under certain conditions, its associated error tends to the minimal error. Assumptions used   are
detailed in Section \ref{stat}. Here, we mention that, in particular,  certain covariance matrices formed from
specifically chosen signals in $K_{_Y}$ and $K_{_X}$ are assumed to be known or estimated.
 Subsequently, all random signals under consideration are finite.

The paper is arranged as follows. After some notations and definitions are introduced in the next subsection, a rigorous
formulation of the problem is given in Section \ref{form}. Some approaches for finding alternate filters are discussed in
Section \ref{disc}. Preliminary results concerning an orthogonalization of random signals are then given in Section
\ref{prel}. The main results of the paper, involving the estimator determination and an error analysis, are presented in
Section \ref{main}. There, it is shown, in Theorems \ref{th2}--\ref{th3} and
Remarks \ref{rem4}--\ref{rem5}, that the proposed filter has several free parameters that can be manipulated to improve
filter performance. They are a number of terms $p$ of the filter $\f$, and sets $S_{_X}$ and  $S_{_Y}$.
Moreover, our filter is determined in terms of pseudo-inverse matrices \cite{bou1} and, therefore, it always exists.

\subsection{Notation. Formulation of the problem}\label{form}

\subsubsection{General notation}\label{not}

For the rigorous statement of the problem, we need the following
notation.

 Let $(\Omega, \Sigma, \mu)$ denote a probability space, where $\Omega = \{\omega\}$ is the set of outcomes,
 $\Sigma$ a $\sigma$--field of measurable subsets in $\Omega$ and $\mu:\Sigma \rightarrow [0,1]$ an
associated probability measure on $\Sigma$.

Suppose that  $\x$ and $\y$ are random signals such that $\x \in L^{2}(\Omega,{\mathbb R}^{m})$ and $\y \in L^{2}
(\Omega,{\mathbb R}^{n})$ where $\x = (\x^{(1)},\ldots, \x^{(m)})^T$ and $\y = (\y^{(1)},\ldots,\y^{(n)})^T$ with
$\x^{(i)},$ $ \y^{(j)}\in L^{2}(\Omega,{\mathbb R})$ for  $i=1,\ldots,m$ and $j=1,$ $\ldots,n$, respectively.

Let
 \begin{eqnarray} \label{xyee1}
\langle\x^{(i)}, \y^{(j)}\rangle =  \int_{\Omega} \x^{(i)}(\omega) \y^{(j)}(\omega) d\mu (\omega)  < \infty
 \end{eqnarray}
 and
 $$
 E_{xy} =E\{\x \y^T\} = \{\langle\x^{(i)}, \y^{(j)}\rangle\}_{i,j=1}^{m,n}\in \rt^{m\times n}.
 $$
We introduce the norm
 $$
\|\x\|^2_{_E} = \int_{\Omega} \|\x(\omega)\|^2_2 d \mu (\omega),
$$
where $\|\x(\omega)\|_2$ is the Euclidean norm of $\x(\omega)$. In the sequel we shall also make reference to the
Frobenius norm $\|A\|$ of matrix $A$.

If $M_{m n}\in\rt^{m\times n}$, then we define an operator $\mathcal M_{m n}: L^{2}(\Omega,{\mathbb R}^{n}) \rightarrow
L^{2}(\Omega,{\mathbb R}^{m})$ by
 \begin{eqnarray} \label{m1}
[\mathcal M_{m n} (\y)](\omega ) = M_{m n}[\y(\omega)].
  \end{eqnarray}
To simplify notation, we omit the index $mn$ for $M_{m n}$ and $\mathcal M_{m n}$ and write (\ref{m1}) as
 \begin{equation}  \label{m}
[\mathcal M (\y)](\omega ) = M[\y(\omega)].
  \end{equation}
Henceforth, an operator defined similarly to that of (\ref{m}) will be denoted by a calligraphic letter.

\subsubsection{Filter structure}

Some more notation should be used to describe the filter under consideration.

The proposed filter $\f$ is specified below in (\ref{ff1}) by means of operators $\q_j$, $\rrr_j$ and $\ttt_j$.
In particular, $\q_1$, $\ldots,$ $\q_{p}$ are used to transform an observable signal $\y\in K_{_Y}$ to $p$ different signals.
By means of the orthogonalization procedure provided in Section \ref{prel}, operators $\rrr_1,\ldots,\rrr_p$ reduce the set
of related matrix equations  to $p$ {independent} equations  (see, below, $(\ref{int2})$ and
$(\ref{tjew})$, respectively).  Their solution requires
much less computational effort than would be required with the full set of equations. Therefore,  operations $\q_j$ and
$\rrr_j$ are used to reduce the overall numerical work needed for the implementation of the proposed filter.
 Operators $\ttt_i,\ldots,\ttt_p$
are determined from interpolation conditions given below by (\ref{int2}) to adjust the  accuracy in signal estimation.

In more detail, the point of using $\q_j$ and $\rrr_j$ is explained in Remark \ref{rem-ort} and Section \ref{choice}.
Here, we introduce $\q_j$ and  $\rrr_j$ as follows.

\subsubsection{Notation related to the filter structure}\label{not}

For  $\q_j:L^{2}(\Omega,{\mathbb R}^{n}) \rightarrow L^{2}(\Omega,{\mathbb R}^{n})$ where  $j=1,\ldots,p$ and $p$
is any positive integer, we denote
 \begin{equation} \label{q}
\vv_j = \q_j (\y)\qa \vv_{jk} = \q_j (\y_k),  
  \end{equation}
where $\y_k$ belongs to the set $S_{_Y}$ introduced above in Section \ref{basic}. For example, if $\y=\y(\cdot,t)$ where $\y(\cdot,t)$ is some arbitrary stationary time series and
$t\in [a, b]\subseteq\rt$ is the time variable, then we can define $\q_1,\ldots,\q_p$ to be
time-shifting operators such that
\begin{equation} \label{uuj}
\vv_1= \y(\cdot,t), \quad \vv_2= \y(\cdot,t-1),\quad \ldots, \quad
\vv_p= \y(\cdot,t-p).
\end{equation}
Such a case has been considered, in particular, in \cite{man1}.

To demonstrate and justify the flexibility of the filter $\f$ in (\ref{ff1}), below, with respect to
the choice of $\q_1,\ldots,\q_p$, we mainly study the case   where $\q_1,\ldots,\q_p$ are arbitrary.
Some other explicit choices of $\q_1,\ldots,\q_p$ are presented in Section \ref{choiceqi} where
we also discuss the benefits associated with some particular forms of $\q_1,\ldots,\q_p$.

Next, for $j=1,\ldots,p,$  let $\rrr_j: L^{2}(\Omega,{\mathbb R}^{n})\rightarrow
L^{2}(\Omega,{\mathbb R}^{n})$ be operators such that, for $k=1,\ldots,p,$  signals
 \begin{equation} \label{z}
\ww_{1k}=\rrr_1 (\vv_{1k}),\quad \ldots, \quad \ww_{p k}=\rrr_{p}(\vv_{p k})
 \end{equation}
are mutually orthogonal in the sense of Definition \ref{def2} given in Section \ref{prel}  below. A method for
determining $\rrr_1$, $\ldots,$ $\rrr_p$ (and consequently, $\ww_{1 k}$, $\ldots,$ $\ww_{p k}$) is also given in Section \ref{prel}. The orthogonalization
is used to reduce the computation of large matrices to a computation of a sequence of much smaller
matrices. Such a procedure leads to the substantial reduction in computational load. See Remark \ref{rem-ort} in
Section \ref{det} for more detail.

We also use the notation
$$
\ww_{1}=\rrr_1 (\vv_{1}),\quad \ldots, \quad \ww_{p }=\rrr_{p}(\vv_{p}).
$$

Since any random signal can be transformed to a signal with zero mean, we shall assume that all signals $\y$
 and $\x$ have zero mean. By a similar argument, we also assume that
signals $\ww_1, \ldots, \ww_p$ have zero mean.

\subsubsection{Statement of the problem}\label{stat}

 Let us assume that
$
K_{_X}\subseteq L^{2}(\Omega,{\mathbb R}^{m})$ and $ K_{_Y}\subseteq L^{2}(\Omega,{\mathbb R}^{n})$, and as before,
$S_{_X}=\{\x_1,\ldots, \x_p\}\subseteq K_{_X} \hspace*{2mm}\mbox{and}\hspace*{2mm} S_{_Y}=\{\y_1,  \ldots,\y_p\}
\subseteq K_{_Y}.
$
In particular, $S_{_X}$ can be an $\varepsilon_{_X}$-set  for $K_{_X}$ and $S_{_Y}$
an $\it{\varepsilon}_{_Y}$-set for $K_{_Y}$.  In this regard, see Definition \ref{def-eps}
in Section \ref{err} below\footnote{More generally, a set $S\subseteq T$ of a Banach space $T$ with a
norm $||\cdot||_T$ is called an $\epsilon$-set for  $T$ if for given $\varepsilon >0$ and any $t\in T$,
there is at least one $s\in S$ such that $||t - s||_T\leq \varepsilon$. The finite $\varepsilon$-set $S$
always exists if $T$ is a compact set \cite{kol1}.}. In Section  \ref{choicesx}, some choice of sets $S_{_X}$
and  $S_{_Y}$ is considered.

For any $\y\in K_{_Y}$, let a filter $\f:K_{_Y}\rightarrow K_{_X}$ be presented in the form
\begin{equation}\label{ff1}
\f(\y) = \sum_{j=1}^{p}\ttt_j(\ww_j)= \sum_{j=1}^{p}\ttt_j\rrr_j\q_j (\y),
\end{equation}
where the $\ttt_j$ is defined by a matrix $T_j\in\rt^{m\times n}$ similarly to (\ref{m}).

{\sf\em The problem} is to find  $T_1,\ldots,T_{p}$ so that
\begin{equation}\label{int2}
\sum_{j=1}^{p}T_j E_{w_{jk}w_{kk}} =E_{x_{k}w_{kk}} \quad \mbox{for
$k=1,\ldots,p$}.
\end{equation}

We note that (\ref{int2}) are  interpolation-like conditions. 
The assumptions we invoke are that

(a) $\x\in K_{_X}$ is unobservable and unknown (i.e. a
constructive analytical form of $\x$ is unknown),

(b) $\y \in K_{_Y}$ is observable but also unknown in the same
sense as for $\x$.
\\

To emphasize the dependence of $\f$, defined by $(\ref{ff1})$,  on $p$ we denote $\f_{(p)}:=\f$.

\begin{definition}\label{def1}
The filter $\f_{(p)}$  is called the {\sf\em interpolation filter of the $p$th order}.
\end{definition}

\subsection{Discussion of the problem}
\label{disc}

The issues related to the statement of the problem are considered in the following Remarks.

\begin{remark}\label{rem2} The sets $K_{_X}$ and $K_{_Y}$ are infinite. Therefore, finding a Wiener-like filter that
minimizes $||\x - \f_{(p)}(\y)||_{_E}$  for every individual pair $\{\x, \y\}\in K_{_X}\times K_{_Y}$, {means that one
needs to find and use an infinite set of filters. Clearly, it } makes no sense { in practice}. For the particular
case when $K_{_X}$ and $K_{_Y}$ are finite sets, i.e. $S_{_X}= K_{_X}$ and $S_{_Y} = K_{_Y}$,  some reasonable approaches
to finding an optimal filter $\f:S_{_Y}\rightarrow S_{_X}$ could be as follows. A possible approach is to find a $\f_{(p)}$ that
minimizes $||\et - \f_{(p)}(\xxi)||^2_{_E}$ where $\et = [\x_1,\x_2,\ldots,\x_p]^T$ and $\xxi = [\y_1,\y_2,\ldots,\y_p]^T$.
A second approach is to find a filter determined from $p$ Wiener-like sub-filters $\f_1,\ldots,\f_p$ that are determined
for each pair $\{\x_k, \y_k\}\in K_{_X}\times K_{_Y}$ with $k=1,\ldots,p$. In both approaches, $\f$ and $\f_k$ (with
$k=1,\ldots,p$) can be chosen, in particular, in the form $(\ref{ff1})$.

Nevertheless, even in such a simplified case (i.e. when $S_{_X}=K_{_X}$ and $S_{_Y}=K_{_Y}$), the first approach would
imply a significant computational effort associated with the computation of large $np\times np$ matrices. The second approach
would require finding a `recognizer' that would recognize a particular sub-filter $\f_k$ from a set of $p$ sub-filters
$\f_1,\ldots,\f_p$, chosen for a particular input signal $\y_k$. Unlike the above two approaches, it will be shown in Section
\ref{det} that the filter satisfying conditions $(\ref{int2})$ requires computation of much lesser $n\times n$ matrices and,
therefore, is more computationally effective. Second, the filter proposed below does not require any `recognizer'. Third,
our filter is applicable to infinite sets $K_{_X}$, $K_{_Y}$.
\end{remark}

\begin{remark}\label{rem3} The parameterized signals considered in Section $\ref{motiv}$,  $\y = \y(\cdot,\alpha)$ and
$\x = \x(\cdot,\alpha)$ for each particular vector of parameters $\alpha\in C^q\subseteq \rt^q$, belong to the particular example
of infinite signal sets discussed above. The Wiener-like filtering approach applied to $\y(\cdot,\alpha)$ and
$\x(\cdot,\alpha)$ leads to finding  ${\displaystyle \min_{\mathcal F} ||\x(\cdot,\alpha) - \f [(\cdot,\alpha)]||^2_{_E}}$ which is a
function of $\alpha$. This means that such a filter $\f$ should be found for each $\alpha$, which is again of little
practical use. An alternative approach is to consider $\y$ and $\x$ as functions $\y: \Omega \times C^q \rightarrow \rt^n$ and
$\x: \Omega \times C^q \rightarrow \rt^m$, respectively, and then to state  a minimization problem in terms of the new norm
$\displaystyle \|\x\|^2_{_{C, E}}=\int_{C^q}\int_{\Omega} \|\x(\omega)\|^2 d \mu (\omega) d \alpha$. Such a problem is
different from that stated above and we do not consider it here.
\end{remark}

\section{Main results} \label{main}


\subsection{Orthogonalization of random signals}\label{prel}

The results presented in this Section will be used in the solution of problem (\ref{ff1})-(\ref{int2})
given in Section \ref{det}.

\begin{definition}\label{def2} The random signals $\ww_{1k},\ww_{1k}\ldots, \ww_{p-1,k}, \ww_{pk}$
are called  mutually orthogonal if, for any $k=1,\ldots,p$,
 \begin{equation}\label{ezjk}
E_{w_{ik}w_{jk}}=\oo\quad \mbox{for $i\neq j$ with $i,j=1,\ldots,p$}.
 \end{equation}
 Here, ${\mathbb O}$ is the zero matrix.
\end{definition}

In this regard, see also \cite{tor20}--\cite{tor103}. The following elementary example illustrates Definition \ref{def2}.
\begin{example}
Let $\ww_{1k}$ and $\ww_{2k}$ be such that $\ww_{1k}(\omega) = \left[ \begin{array}{r}
\omega \\
 -\omega\end{array}  \right]$ and  $\ww_{2k}(\omega)$ ${\displaystyle  = \left[ \begin{array}{r}
\frac{3}{4}\omega-\omega^2 \\
 -\frac{3}{5}\omega+\omega^3\end{array}  \right]}$ where  $\omega \in\Omega = [0 \hspace{2mm} 1]$.
Then $E_{w_{1k}w_{2k}}=\oo$.
\end{example}

We write $M^\dag$ for the Moore-Penrose pseudo-inverse \cite{bou1} of the matrix $M$  and define
$\kk_{i\ell}:L^{2}(\Omega,{\mathbb R}^{n})\rightarrow L^{2}(\Omega,{\mathbb R}^{n})$  by
\begin{equation}\label{kil}
\label{a} K_{i\ell} = E_{v_{i\ell} w_{\ell k}} E^\dag_{w_{\ell k} w_{\ell k}} + M_{i\ell}(I -
E_{w_{\ell k} w_{\ell k}} E^\dag_{w_{\ell k} w_{\ell k}})
\end{equation}
with $M_{i\ell}\in \rt^{n\times n}$ arbitrary and $i,\ell = 1,\ldots, p$..

The symbol $\tth$ denotes the zero vector.

\begin{lemma} \label{ort3} Let $\g_{jk}: L^{2}(\Omega,{\mathbb R}^{n})\rightarrow
L^{2}(\Omega,{\mathbb R}^{n})$ be a linear continuous  operator for $j,k=1,\ldots, p$. Let
random signals $\vv_{1k},\ldots, \vv_{p-1,k}, \vv_{pk}$ be such that, for any $k=1,\ldots,p$,
 \begin{equation}\label{gujk}
\g_{1k}(\vv_{1k}) + \ldots + \g_{p-1, k}(\vv_{p-1, k})+ \g_{p k}(\vv_{p k}) = \tth
 \end{equation}
 if and only if $\g_{1 k},\ldots,\g_{p-1, k}, \g_{p k}$ are the zero operators.

 Then, for $k=1,\ldots, p$, random signals $\ww_{1k}=\rrr_1 (\vv_{1k}),\ldots, \ww_{pk}=\rrr_p (\vv_{p k})$ with
 $\rrr_1$, $\ldots$, $\rrr_p$ such that
\begin{equation}\label{f1}
\ww_{1k} = \vv_{1k}\quad\mbox{and, for $i=2,\ldots, p,$}\quad
\ww_{ik} = \vv_{ik} -\sum_{l=1}^{i-1}\kk_{il}(\ww_{lk})
\end{equation}
are mutually orthogonal.
\end{lemma}

\noindent
{\bf Proof.}  We wish for (\ref{ezjk}) to be true. If $K_{i\ell}$ has been chosen so that  condition (\ref{ezjk}) is
true   for all $\ww_{\ell k}$ with $\ell=1,\ldots,i-1$ and $k=1,\ldots, p$, then we have (see (\ref{xyee1}))
\begin{equation}\label{ekw1}
E\left\{\left[\vv_{ik} -\sum_{l=1}^{i-1}\kk_{il}(\ww_{lk})\right]\ww^T_{jk}\right\} =\oo\quad \Rightarrow \quad
E_{v_{i\ell} w_{\ell k}} - K_{i\ell}E_{w_{\ell k} w_{\ell k}} = \oo.
\end{equation}
Equation (\ref{ekw1}) has a solution \cite{bou1} if and only if
$E_{v_{i\ell} w_{\ell k}} = E_{v_{i\ell} w_{\ell k}} E_{w_{\ell k} w_{\ell k}}^\dag
E_{w_{\ell k} w_{\ell k}}$. The latter is true by \cite{tor100}, p. 168. Then the general solution  \cite{bou1} is given
by (\ref{kil}). Therefore, $\ww_{ik}$ defined by (\ref{f1}) also satisfies the condition (\ref{ezjk}).
Thus, signals $\ww_{1k},\ldots, \ww_{p,k-1}, \ww_{pk}$ are mutually orthogonal for any
$k=1,\ldots,p$.  $\hfill \rule{3mm}{3mm}$

\begin{remark} In  Lemma \ref{ort3}, {\em (\ref{f1})} implies  $\ww_{ik}=\tth$ if $\vv_{ik}$ $=\sum_{l=1}^{i-1}\kk_{il}(\ww_{lk})$.
At the same time, we are interested in non-zero signals $\ww_{1k},\ldots, \ww_{pk}$. That is why, in Lemma \ref{ort3}, the
restriction of the joint independency {\em (\ref{gujk})}  of signals $\vv_{1k},\ldots, \vv_{pk}$ has been imposed.

We also note that {\em (\ref{gujk})} is `an operator version' of the definition of the vector linear independence.
\end{remark}

Below we give exact formulae (\ref{ti}) for the matrices $T_1$, $\ldots,$ $T_p$ that satisfy condition (\ref{int2})
and present an error analysis of the associated estimator  defined by  (\ref{ff1}) and (\ref{ti}).

\subsection{Determination of $T_j$} \label{det}

\begin{theorem}\label{th1} Let $\{\ww_{jk}\}_{j,k = 1}^p$ be a set of
orthogonal random signals defined by (\ref{kil}) and (\ref{f1}).
Matrices $T_1$, $\ldots,$ $T_p$ that satisfy condition
$(\ref{int2})$ are given by
\begin{equation}\label{ti}
T_j = E_{x_j w_{jj}}E^\dag_{w_{jj} w_{jj}} + A_j(I - E_{w_{jj} w_{jj}}E^\dag_{w_{jj}w_{jj}}),
\end{equation}
where the $A_j$ for $j=1,\ldots,p$ are arbitrary matrices.
\end{theorem}

\noindent {\bf Proof.} The proof follows directly from
(\ref{int2}) and (\ref{ezjk}). Indeed, (\ref{ezjk}) implies that
(\ref{int2}) reduces to
\begin{equation}\label{tjew} T_j
E_{w_{jj}w_{jj}}=E_{x_{j}w_{jj}} \quad \mbox{where $j =
1,\ldots,p$.}
\end{equation}
The solution to this equation is given by (\ref{ti}) if and only
if \cite{bou1}
$$
E_{x_j w_{jj}}E^\dag_{w_{jj} w_{jj}}E_{w_{jj} w_{jj}} = E_{x_{j} w_{jj}}.
$$
By \cite{tor100} (p. 168), the last identity is true.  Thus,
(\ref{ti}) is also true.
 $\hfill \rule{3mm}{3mm}$
\bigskip

We note that the matrix $T_j$ is not unique due to the arbitrary matrices $A_j$. In particular, the $A_j$ can be set
identically to be the zero matrix. Such a choice is normally done in the known Wiener-type filters
\cite{fom1,man1,mat1,sch1,tor21,tor20,tor100}.

\begin{remark}\label{rem-ort}
Theorem $1$ and its proof motivate the use of the orthogonalizing operators $\rrr_1, \ldots, \rrr_p$. If the random
signals $\ww_{1k},\ldots,\ww_{p-1,k},\ww_{p k}$ were not orthogonal, then condition $(\ref{int2})$ would represent a set
of matrix equations for the $T_1,\ldots,T_p$. The orthogonalization of the signals $\{\ww_{ik}\}_{i=1}^p$ by Lemma 1
reduces the set of matrix equations $(\ref{int2})$ to $p$ {\sf\em independent} equations $(\ref{tjew})$.  Their solution requires
much less computational effort than would be required with the full set of equations. Although the orthogonalization
procedure by Lemma 1 requires additional computational work, the use of orthogonalizing operators $\rrr_1, \ldots, \rrr_p$
leads to a substantial reduction in the overall computational load needed for the implementation of the proposed filter.
\end{remark}

\begin{remark}\label{ps-inv}
The proposed filter is determined in terms of the pseudo-inverse matrices, therefore, the filter  always exists.

\end{remark}

\subsection{Error analysis} \label{err}

Let us now estimate the error $\|\x - \f_{(p)}(\y)\|^2_{_E}$ associated
with the interpolation filter of the $p$th order $\f_{(p)}$ presented by (\ref{ff1}) and (\ref{ti}).

\begin{definition}  \label{def-eps} $\cite{kol1}$ The set $\{\x_k\}_{k=1}^p$ is an
$\varepsilon_{_X}$-net for $K_{_X}$ if, for any  $\varepsilon_{_X}
\geq 0 $ and $\x\in K_{_X}$, there exists at least one $\x_k$ such
that
$$
\|\x - \x_k\|_{_E}^2\leq \varepsilon_{_X} \quad \mbox{for}\quad
k=1,\ldots,m.
$$
\end{definition}

{  The  $\varepsilon_{_Y}$-net for $K_{_Y}$, $\{\y_j\}_{j=1}^p$, is defined similarly.}

In Theorem 2 below, we show that the error $\|\x -
\f_{(p)}(\y)\|^2_{_E}$ associated with the proposed filter $\f_{(p)}$ for
an {\sf\em infinite set} of signals is asymptotically close, in
the sense $\varepsilon_{_X}, \varepsilon_{_Y} \rightarrow 0$, to the error
associated with the optimal filter, denoted $\p_p$, of
a similar structure developed for an {\sf\em individual} input
signal only. The filter $\p_p$ has been studied in
\cite{tor20} (Section 5.2.2). The filter $\p_p$ is
optimal in the sense of minimizing the associated error for an
{\sf\em individual} input signal. That filter generalizes known
filters developed from the Wiener approach. The error $\|\x -
\p_p(\y)\|^2_{_E}$ associated with the filter
$\p_p$ is given by \cite{tor20}\footnote{$A^{1/2}$ is
defined by the condition $A^{1/2}A^{1/2}=A$.}
\begin{eqnarray}\label{er10}
\|\x - \p_p(\y)\|^2_{_E} = \|E_{xx}^{1/2}\|^2
-\sum_{j=1}^p  \|E_{xw_j}(E_{w_j w_j}^{1/2})^{\dag} \|^2.
\end{eqnarray}

\begin{definition}
We say that a filter $\f_{(p)}$ is asymptotically optimal if its
associated error $\|\x - \f_{(p)}(\y)\|^2_{_E}$ tends to the right hand side of $(\ref{er10})$ as
$\varepsilon_{_X}, \varepsilon_{_Y}  \rightarrow 0$.
\end{definition}

The following theorem, Theorem 2, characterizes the error $\|\x
-\f_{(p)}(\y)\|^2_{_E}$ associated with the proposed filter $\f_{(p)}$ in
terms of the error associated with the optimal filter
(\ref{er10}). In this theorem and in Theorem 3 below, operators
$\q_k$ are arbitrary and operators $\rrr_k$ are defined by Lemma
1.

\begin{theorem}\label{th2} Let $K_{_X}$ and $K_{_Y}$ be compact sets. Let
$S_{_X}=\{\x_k\}_{k=1}^p$  be an $\varepsilon_{_X}$-net for
$K_{_X}$ and $S_{_Y}=\{\y_k\}_{k=1}^p$  an $\varepsilon_{_Y}$-net
for $K_{_Y}$. Then
\begin{eqnarray}\label{err2}
\|\x - \f_{(p)}(\y)\|^2_{_E} \rightarrow \|\x - \p_p(\y)\|^2_{_E}
 \quad \mbox{as} \quad \varepsilon_{_X}, \varepsilon_{_Y}  \rightarrow 0.
\end{eqnarray}
That is, the interpolation filter of the $p$th order $\f_{(p)}$, given by $(\ref{ff1})$, $(\ref{f1})$
and $(\ref{ti})$, is asymptotically optimal.\footnote{Note that we have assumed the two $\varepsilon-$nets possess
the same number of elements, $p$. More general situations can be made the
subject of a future investigation. Note also that, in general,
$p=p(\varepsilon)$.}
\end{theorem}

\noindent
The proof follows the proof of Theorem 3 below.


{ To formulate and prove Theorem 3, we need the following notation}.

Let  $\zz_k:=\ww_{kk}$. Let $E^\dag_{y y}$ and $\q_k(\y)$ satisfy the Lipschitz conditions
\begin{eqnarray}\label{lip1}
\hspace*{-9mm}\|E^\dag_{z_k z_k} - E_{w_kw_k}^{\dag}\|^2 \leq  \lambda_{_E}
\|\zz_k - \ww_{k}\|_{_E}^2, \hspace*{1mm} \|\q_k(\y_{k}) -
\q_k(\y)]\|_{_E}^2\leq  \lambda_{_Q} \|\y_k - \y\|_{_E}^2
\end{eqnarray}
with the Lipschitz constants $\lambda_{_E}$ and $\lambda_{_Q}$,
respectively. Let the operator $\rrr_k$ be bounded so that, for
some $\hat{R_k}>0$,
\begin{eqnarray}\label{rk1}
\|\rrr_k\|_{_O}^2\leq \hat{R_k},
\end{eqnarray}
where $\|\cdot\|_{_O}$ is the operator norm \cite{kol1}.  Let us also denote
\begin{eqnarray*}\label{dk1}
&& D_k = \lambda_{_E}\lambda_{_Q} \hat{R_k}\left[\|\x\|^2_{_E}E_k
+ \|E_{x_k w_k}\|^2\right ] \|E^{1/2}_{w_k w_k}\|^2, \\
&&C = \sum_{k=1}^{p}
\|A_k(I - E_{w_k w_k}E^\dag_{w_k w_k})\|^2 \|E_{w_kw_k}^{1/2}\|^2,\\
&&J_0 = \|\x - \p_p(\y)\|^2_{_E} \qa E_k =\|(E_{w_kw_k}^{1/2})^{\dag}\|^2.
\end{eqnarray*}
As before,  $\ww_k =\rrr_k \q_k (\y) $.

\begin{theorem}\label{th3}
Let $K_{_X}$ and $K_{_Y}$ be compact sets. Let
$S_{_X}=\{\x_k\}_{k=1}^p$  be an $\varepsilon_{_X}$-net for
$K_{_X}$ and $S_{_Y}=\{\y_k\}_{k=1}^p$  an $\varepsilon_{_Y}$-net
for $K_{_Y}$.  Then an
estimate of the error $\|\x - \f_{(p)}(\y)\|^2_{_E}$ associated with the interpolation filter of the $p$th order
 $\f_{(p)}: K_{_X}\rightarrow K_{_Y}$ presented by $(\ref{ff1})$ and $(\ref{ti})$ is given by
 \begin{equation}\label{er1}
\|\x - \f_{(p)}(\y)\|^2_{_E} \leq \|\x - \p_p(\y)\|^2_{_E} + \sum_{k=1}^{p} \left
(\varepsilon_{_X} \|\ww_k\|^2_{_E}E_k + \varepsilon_{_Y} D_k \right) +C.
 \end{equation}
\end{theorem}

\noindent
 {\bf Proof.} { Since $K_{_X}$ and $K_{_Y}$ are compact sets, a finite $\varepsilon_{_X}$-net and finite
 $\varepsilon_{_Y}$-net,  for $K_{_X}$ and $K_{_Y}$, respectively, always exists \cite{kol1}.} For any $T_1,\ldots, T_p$, we have
\begin{eqnarray}\label{er2}
\|\x - \f_{(p)}(\y)\|^2_{_E} = \|\x -
\sum_{j=1}^{p}\ttt_j(\ww_j)\|^2_{_E} = J_0 + J_1,
\end{eqnarray}
where $J_0$ is as above and
\begin{eqnarray}\label{j01}
  J_1 =\sum_{k=1}^p \|T_kE_{w_kw_k}^{1/2} - E_{xw_k}(E_{w_kw_k}^{1/2})^{\dag}\|^2.
\end{eqnarray}
The error representation in (\ref{er2}) and (\ref{j01}) follows
from \cite{tor20} under the zero-mean assumption $E[\x] = E[\ww_j]
=\oo$ which we have made in Section \ref{form}.

Next, for $T_j$ given by (\ref{ti}) with $\zz_k:=\ww_{kk}$, the summand of  $J_1$ is
represented as follows:
\begin{eqnarray}\label{er3}
&&\hspace*{-10mm}\|T_kE_{w_kw_k}^{1/2} - E_{xw_k}(E_{w_kw_k}^{1/2})^{\dag}\|^2\nonumber\\
&=&\hspace*{-1mm}\|E_{x_k z_k}E^\dag_{z_k z_k} E_{w_kw_k}^{1/2} -
E_{xw_k}(E_{w_kw_k}^{1/2})^{\dag} + A_k(I - E_{z_k z_k}E^\dag_{z_k
z_k}) E_{w_kw_k}^{1/2}\|^2
\end{eqnarray}
where
\begin{eqnarray}\label{er4}
&&\hspace*{-10mm}E_{x_k z_k}E^\dag_{z_k z_k} E_{w_kw_k}^{1/2}
- E_{xw_k}(E_{w_kw_k}^{1/2})^{\dag}\nonumber\\
& = &(E_{x_k z_k}E^\dag_{z_k z_k}
-E_{xw_k}E_{w_kw_k}^{\dag})E_{w_kw_k}^{1/2} \nonumber\\
& = & [E_{x_k z_k}(E^\dag_{z_k z_k} - E_{w_kw_k}^{\dag}) + (E_{x_k
z_k} -
E_{xw_k})E_{w_kw_k}^{\dag}]E_{w_kw_k}^{1/2}\nonumber\\\label{er41}
& = & [E_{x_k z_k}(E^\dag_{z_k z_k} - E_{w_kw_k}^{\dag}) + (E_{x_k z_k} - E_{xz_k})E_{w_kw_k}^{\dag}\\
&&\hspace*{65mm}+ (E_{x z_k} - E_{xw_{k}})E_{w_kw_k}^{\dag}]E_{w_kw_k}^{1/2}
\end{eqnarray}
because of the relation $(E_{w_kw_k}^{1/2})^{\dag} =E_{w_kw_k}^{\dag}E_{w_kw_k}^{1/2}$ \cite{tor100}.
In (\ref{er41}),
\begin{eqnarray*}\label{er5}
\|E^\dag_{z_k z_k} - E_{w_kw_k}^{\dag}\|^2 &\leq & \lambda_{_E}
\|\zz_k - \ww_{k}\|_{_E}^2
\leq \lambda_{_E} \|\rrr_k [\q_k(\y_{k}) - \q_k(\y)]\|_{_E}^2\nonumber\\
&\leq &\lambda_{_E}  \lambda_{_Q} \hat{R_k}\|\y_{k} - \y\|_{_E}^2
\leq \varepsilon_{_Y}\lambda_{_E}  \lambda_{_Q} \hat{R_k},
\end{eqnarray*}
and on the basis of the Cauchy--Schwarz inequality,
\begin{eqnarray*}\label{er6}
\|E_{x_k z_k} - E_{xz_k}\|^2 &=& \left\| \left  \{\int_\Omega
[\x^{(i)}_k(\omega) -
\x^{(i)}(\omega)] \zz^{(j)}_k(\omega) d \mu (\omega) \right \}_{i,j=1}^{m, n} \right\|^2\nonumber\\
&=& \sum_{i=1}^m \sum_{j=1}^n \left [\int_\Omega
[\x^{(i)}_k(\omega) -
\x^{(i)}(\omega)] \zz^{(j)}_k(\omega) d \mu (\omega)\right ]^2\nonumber\\
&\leq &  \sum_{i=1}^m\sum_{j=1}^n\int_\Omega
[\x^{(i)}_k(\omega)-\x^{(i)}(\omega)]^2 d \mu
(\omega) \int_\Omega [\zz^{(j)}_k(\omega)]^2 d \mu (\omega)\nonumber\\
&=&  \int_\Omega \sum_{i=1}^m
[\x^{(i)}_k(\omega)-\x^{(i)}(\omega)]^2 d \mu
(\omega) \int_\Omega \sum_{j=1}^n[\zz^{(j)}_k(\omega)]^2 d \mu (\omega)\nonumber\\
&= &  \|\x_k - \x\|^2_{_E}  \|\zz_k\|^2_{_E}\nonumber\\
&\leq & \varepsilon_{_X} \|\zz_k\|^2_{_E}
\end{eqnarray*}
and
\begin{eqnarray*}\label{er7}
\|E_{x z_k} - E_{xw_{k}}\|^2 &=& \left \| \left  \{\int_\Omega
\x^{(i)}(\omega) [\zz^{(j)}_k(\omega) - \ww^{(j)}_k(\omega)]  d
\mu (\omega) \right \}_{i,j=1}^{m,n}
 \right \|^2\nonumber\\
&=& \sum_{i=1}^m \sum_{j=1}^n \left [\int_\Omega \x^{(i)}(\omega)
[\zz^{(j)}_k(\omega) - \ww^{(j)}_k(\omega)] d \mu (\omega)\right ]^2\nonumber\\
&\leq &  \sum_{i=1}^m \sum_{j=1}^n \int_\Omega
[\x^{(i)}(\omega)]^2 d \mu (\omega) \int_\Omega
[\zz^{(j)}_k(\omega) -\ww^{(j)}_k(\omega)]^2
d \mu (\omega)\nonumber\\
&= &  \int_\Omega \sum_{i=1}^m [\x^{(i)}(\omega)]^2 d \mu (\omega)
\int_\Omega \sum_{j=1}^n[\zz^{(j)}_k(\omega)
-\ww^{(j)}_k(\omega)]^2
d \mu (\omega)\nonumber\\
&=& \|\x\|^2_{_E} \|\zz_k - \ww_k\|^2_{_E}\nonumber\\
&\leq & \varepsilon_{_Y}\lambda_{_E}  \lambda_{_Q}
\hat{R_k}\|\x\|^2_{_E}.
\end{eqnarray*}
Therefore,
\begin{eqnarray*}\label{er8}
&&\hspace*{-10mm}\|T_kE_{w_kw_k}^{1/2} - E_{xw_k}(E_{w_kw_k}^{1/2})^{\dag}\|^2\\
&\leq &\|E_{x_k z_k}E^\dag_{z_k z_k} E_{w_kw_k}^{1/2}
- E_{xw_k}(E_{w_kw_k}^{1/2})^{\dag}\|^2 + \|A_k(I - E_{z_k z_k}E^\dag_{z_k z_k})E_{w_kw_k}^{1/2}\|^2\\
&&\hspace{5mm}\leq  (\varepsilon_{_Y}\lambda_{_E}  \lambda_{_Q} \hat{R_k}\|E_{x_k w_k}\|^2\|E^{1/2}_{w_k w_k}\|^2
+ \varepsilon_{_X} \|\zz_k\|^2_{_E}E_k\|E^{1/2}_{w_k w_k}\|^2 \\
&& \hspace{9mm}+\varepsilon_{_Y}\lambda_{_E}  \lambda_{_Q}\hat{R_k}\|\x\|^2_{_E}E_k \|E^{1/2}_{w_k w_k}\|^2
 + \|A_k(I - E_{z_k z_k}E^\dag_{z_k z_k})\|^2 \| E_{w_kw_k}^{1/2}\|^2
\end{eqnarray*}
and
\begin{eqnarray*}\label{er9}
&&\hspace{-7mm}\|\x - \f_{(p)}(\y)\|^2_{_E} \\
&& \hspace{0mm}\leq J_0 + \sum_{k=1}^{p}
\left \{\left(\varepsilon_{_X} \|\zz_k\|^2_{_E}E_k +
\varepsilon_{_Y}\lambda_{_E}\lambda_{_Q} \hat{R_k}\left[\|\x\|^2_{_E}E_k
+ \|E_{x_k w_k}\|^2\right ]\right)
\right.\\
&& \hspace{50mm}\left . + \|A_k(I - E_{w_k w_k}E^\dag_{w_k w_k})\|^2  \right \} \|E_{w_kw_k}^{1/2}\|^2.
\end{eqnarray*}
The theorem is proved.  $\hfill \rule{3mm}{3mm}$


\noindent
{\bf Proof of Theorem 2.} We have $I - E_{z_k z_k}E^\dag_{z_k z_k} = I - E^{1/2}_{z_k z_k}(E^\dag)^{1/2}_{z_k z_k}$
\cite{tor20}. Therefore,
$$
(I - E_{z_k z_k}E^\dag_{z_k z_k})E_{w_kw_k}^{1/2} =
E_{w_kw_k}^{1/2} - E^{1/2}_{z_k z_k}(E^\dag_{z_k z_k})^{1/2}
E_{w_kw_k}^{1/2}.
$$
Thus, $A_k(I - E_{z_k z_k}E^\dag_{z_k z_k}) E_{w_kw_k}^{1/2}
\rightarrow 0$ as $\varepsilon_{_Y} \rightarrow 0$. Then, the
statement of Theorem 2 follows from (\ref{er1}). $\hfill\rule{3mm}{3mm}$

\begin{remark}\label{rem4} It follows from $(\ref{err2})$ and $(\ref{er1})$ that
the error decreases if $p$ increases.
\end{remark}

\begin{remark} \label{rem5} A direct comparison of $(\ref{err2})$ and $(\ref{er1})$ shows that
a proper choice of sets $S_{_X}$ and $S_{_Y}$ leads to a decrease in the error
$\|\x - \f_{(p)}(\y)\|^2_{_E}$. { In other words, the error decreases when  $\varepsilon_{_Y}$ and $\varepsilon_{_X}$ become
smaller. The latter, of course, implies an increase in computational work because smaller $\varepsilon_{_Y}$ and
$\varepsilon_{_X}$ are, in particular, achieved by an increase in the number of members of  $S_{_X}$ and $S_{_Y}$.
Thus, in practice, a tradeoff between the accuracy of estimation  and an
associated computational price should be sought.}
\end{remark}

Thus, the free parameters for the proposed  filter presented by $(\ref{ff1})$ and $(\ref{ti})$ are its
number of terms $p$, and sets $S_{_X}$ and  $S_{_Y}$.  { In particular, it is interesting to consider the case when
sets $S_{_X}$ and  $S_{_Y}$ coincide with sets $K_{_X}$ and  $K_{_Y}$, respectively.  The next Corollary 1 establishes
an error estimate associated with this case. A direct comparison of the error estimates (\ref{er1}) and (\ref{er21})
below shows that the error of the filter decreases if $K_{_X}=S_{_X}$ and $K_{_Y}=S_{_Y}$.

\begin{corollary} Let $K_{_X}=S_{_X}=\{\x_1,\ldots, \x_p\}$ and $K_{_Y}=S_{_Y}=\{\y_1,\ldots, \y_p\}$. Then, for any $\x_k \in K_{_X}$ and $\y_k \in K_{_Y}$, an
estimate of the error $\|\x_k - \f_{(p)}(\y_k)\|^2_{_E}$ associated with the
filter $\f_{(p)}: K_{_X}\rightarrow K_{_Y}$ presented by $(\ref{ff1})$ and $(\ref{ti})$, is given by
 \begin{equation}\label{er21}
\hspace*{-4mm}\|\x_k - \f_{(p)}(\y_k)\|^2_{_E} = \|\x_k - \p_p(\y_k)\|^2_{_E} + \sum_{j=1}^{p}
\|[E_{x_jw_j} - E_{x_kw_j}](E_{w_jw_j}^{1/2})^{\dag}\|^2.
 \end{equation}

\end{corollary}

\noindent
 {\bf Proof.} For $K_{_X}=S_{_X}$ and $K_{_Y}=S_{_Y}$, we have $\vv_j = {\vv}_{jk}$, $\ww_j=\ww_{jk}$,
 $\ww_{kk}=\zz_k=\ww_k$ and, therefore,
$T_k = E_{x_k w_k}E^\dag_{w_k w_k} + A_k(I - E_{w_k w_k}E^\dag_{w_k w_k})$. Thus, (\ref{er3}) is represented as follows:
\begin{eqnarray}\label{er31}
&&\hspace*{-17mm}\|T_j E_{w_jw_j}^{1/2} - E_{x_kw_j}(E_{w_jw_j}^{1/2})^{\dag}\|^2\nonumber\\
&=&\|E_{x_j w_j}E^\dag_{w_j w_j} E_{w_jw_j}^{1/2} - E_{x_k w_j}(E_{w_jw_j}^{1/2})^{\dag}
+ A_j(I - E_{w_j w_j}E^\dag_{w_jw_j}) E_{w_jw_j}^{1/2}\|^2,
\end{eqnarray}
where $\x$ is replaced with $\x_k$. Since $(E_{w_jw_j}^{1/2})^{\dag} =E_{w_jw_j}^{\dag}E_{w_jw_j}^{1/2}$ and
$I - E_{w_j w_j}E^\dag_{w_j w_j} = I - E^{1/2}_{w_j w_j}(E^\dag)^{1/2}_{w_j w_j}$ (see above), we have
$$
(I - E_{w_j w_j}E^\dag_{w_jw_j}) E_{w_jw_j}^{1/2} = \oo
$$
and
\begin{eqnarray}\label{tjezj}
\|T_j E_{w_jw_j}^{1/2} - E_{x_kw_j}(E_{w_jw_j}^{1/2})^{\dag}\|^2
= \|[E_{x_jw_j}- E_{x_kw_j}](E_{w_jw_j}^{1/2})^{\dag}\|^2.
\end{eqnarray}
Here, $\oo$ is the zero matrix. Then  (\ref{er21}) follows from (\ref{er2})--(\ref{er3}) with  $\x=\x_k$, and
(\ref{er31})--(\ref{tjezj}).  $\hfill\rule{3mm}{3mm}$
}

\subsection{Choice of operators $Q_i$, and sets  $S_{_X}$ and  $S_{_Y}$}
\label{choice}

\subsubsection{Choice of $Q_i$}
\label{choiceqi}

In the filter model (\ref{ff1}), operators $\q_1$, $\ldots,$
$\q_{p}$ are used to transform observable signals $\y$ to $p$
signals $\vv_1$, $\ldots$,$\vv_{p}$ in accordance with (\ref{q}).
In particular, the $\vv_1$, $\ldots$,$\vv_{p}$ may coincide with
the $\vv_{1k}$,  $\ldots$,$\vv_{pk}$ -- see (\ref{q}). However,
our model requires that the signals $\vv_1$, $\ldots$,$\vv_{p}$ be
{\em distinct}. If some of the $\vv_1$, $\ldots$,$\vv_{p}$ were
the same, say $\vv_{i}=\vv_{j}$ with $i\neq j$ for some
$i,j=1,\ldots,p$, and if $\ttt_i$, $\rrr_i$, $\q_i$ were fixed,
then terms in (\ref{ff1}) associate with $\vv_{i}$ and $\vv_{j}$,
would coincide. In such a case, the number of terms in the
filter (\ref{ff1}) would be less than $p$. However, we wish to
have the number of terms $p$ fixed because each term in
(\ref{ff1}) contributes to the filter performance. Hence, to
keep the structure (\ref{ff1}), the signals  $\vv_1,
\ldots$,$\vv_{p}$ need to be distinct.

Here, we consider some examples for choosing the $\q_i$.

(i) Let $\Delta^g$ be a subset in $\rt^g$ and let
$\{\alpha_1,\ldots,\alpha_{p}\}\subset \Delta^g$ be a set of given
signals  with
 $\alpha_i\neq \alpha_j$ for $i\neq j$.
For any $\alpha \in \rt^g$ and for $i=1,\ldots,p,$ let us choose
$\q_j:L^{2}(\Omega\times \Delta^g,{\mathbb R}^{n}) \rightarrow
L^{2}(\Omega\times \Delta^g,{\mathbb R}^{n})$ so that
\begin{equation}\label{u1}
\hspace*{-7mm}\vv_{1}(\cdot,\alpha)=\y(\cdot,\alpha), \hspace*{2mm}
\vv_{2}(\cdot,\alpha)=\y(\cdot,\alpha-\alpha_1), \hspace*{2mm} \ldots,
\hspace*{2mm} \vv_{{p}}(\cdot,\alpha)=\y(\cdot,\alpha-\alpha_{p}).
\end{equation}
Such a choice of $\q_i$ is motivated by a generalization of the
case considered in \cite{man1}.

(ii) Let $\Delta^g (\alpha)=[0, \beta^{(1)}]\times \ldots \times
[0, \beta^{(g)}] \subseteq\Delta^g $ with $\beta^{(i)}>0$ for all
$i=1,\ldots,g.$ Let $\alpha = [\alpha^{(1)},\ldots,\alpha^{(g)}]
\in \Delta^g (\alpha)$. For $i=1,\ldots,p$ with $p\leq g+1$, and
$r_{1}(\alpha),\ldots, r_{p}(\alpha)$ known functions, we set
$\q_j:L^{2}(\Omega\times \Delta^g(\alpha),{\mathbb R}^{n})
\rightarrow L^{2}(\Omega\times \Delta^g(\alpha),{\mathbb R}^{n})$
so that
\begin{eqnarray}\label{u2}
&&\vv_{0}(\cdot,\alpha)=\y(\cdot,\alpha), \quad
\vv_{1}(\cdot,\alpha) =\int_0^{\alpha^{(1)}}
r_1(\alpha) \y(\cdot,\alpha) d\alpha^{(1)}, \ldots,\\
&&\vv_{{p}}(\cdot,\alpha)=\int_0^{\alpha^{(p)}}r_{p}(\alpha)
\y(\cdot,\alpha) d\alpha^{(p)}.
\end{eqnarray}

(iii) For $\alpha_{(1)},$ $\ldots,$ $\alpha_{(p)}\in \Delta^g
(\alpha)$ and $r_{1}(\alpha_{(1)}),$ $\ldots,$
$r_{p}(\alpha_{(p)})$ known functions, let us put $\q_i$ so that
\begin{equation}\label{u31}
\hspace{-9mm} \vv_{0}(\cdot,\alpha)=\y(\cdot,\alpha), \quad
\vv_{1}(\cdot,\alpha) =\int_{\Delta^g (\alpha)}r_{1}(\alpha_{(1)})
\y(\cdot,\alpha_{(1)}) d\alpha_{(1)},
\end{equation}
\begin{equation}\label{u32}
\hspace{-9mm}\vv_{2}(\cdot,\alpha)=\int_{\Delta^g (\alpha)}\int_{\Delta^g
(\alpha)}r_{1}(\alpha_{(1)})r_{2}(\alpha_{(2)})
  \y(\cdot,\alpha_{(1)})  \y(\cdot,\alpha_{(2)})  d\alpha_{(1)} d\alpha_{(2)},\quad \ldots,
\end{equation}
\begin{equation}\label{u33}
\hspace{-9mm}\vv_{p}(\cdot,\alpha)=\hspace*{-2mm}\underbrace{\int_{\Delta^g
(\alpha)}\hspace*{-2mm}\ldots \hspace*{-2mm}\int_{\Delta^g (\alpha)} }_{p}
r_{1}(\alpha_{(1)})\ldots
r_{p}(\alpha_{(p)})\y(\cdot,\alpha_{(1)}) \ldots
\y(\cdot,\alpha_{(p)}) d\alpha_{(1)}\ldots d\alpha_{(p)}.
\end{equation}

Choosing $\q_i$ as in the form (\ref{u2}) and (\ref{u31})--(\ref{u33}) is motivated by  the
following observation. It is clear, that the lesser the difference $\|\x - \f_{(p)}(\y)\|^2_{_E}
=\|\x - \sum_{j=1}^{p}\ttt_j\rrr_j\q_j (\y)\|^2_{_E}$,  the better the estimation of $\x$ by the
filter $\f_{(p)}(\y)= \sum_{j=1}^{p}\ttt_j\rrr_j\q_j (\y)$. Here, $\vv_j=\q_j (\y)$. In turn, the
closer $\vv_j$ is to $\x$, the smaller the value of  $\|\x - \f_{(p)}(\y)\|^2_{_E}$. For example, if $\x=\vv_j$ for all
$j=1,\ldots,p$ then one can set $p=1$, choose $\ttt_1$ and $\rrr_1$ equal to the identity, and
then $\|\x - \f_{(p)}(\y)\|^2_{_E}=0$. Therefore, it is reasonable to choose $\vv_j$
as an approximation to $\x$. The Volterra series is one possible approximation to $\x$ and the
latter choices for $\q_i$ are associated with the Volterra series \cite{sch1}.

In general, $\q_j$  might be chosen so that  $\vv_{j}=\q_j(\y)$, with $j=1,\ldots,p$, are
approximations to $\x$ constructed from the known methods. Such methods are
 considered, in particular, in \cite{fer1}, \cite{gol1},\cite{kol1}--\cite{mat1}, \cite{tor21}--\cite{tor103}.

\subsubsection{Choice of sets $S_{_X}$ and  $S_{_Y}$} \label{choicesx}

By Theorems 2 and 3, $S_{_X}=\{\x_k\}_{k=1}^p$ is an
$\varepsilon_{_X}$-net for $K_{_X}$ and $S_{_Y}=\{\y_k\}_{k=1}^p$
is an $\varepsilon_{_Y}$-net for $K_{_Y}$. It follows from
Theorems 2 and 3 that the error associated with the proposed
filter decreases when $\varepsilon_{_X}$ and $\varepsilon_{_Y}$
decrease.

In practice, signals $\x\in K_{_X}$ and $\y\in K_{_Y}$ are
parameterized similarly to those considered  in Section
\ref{choiceqi} above, i.e. $\x =\x(\cdot,\alpha)$ and $\y
=\y(\cdot,\alpha)$ where $\alpha \in \Delta^g (\alpha).$ In such a
case, the sets $S_{_X}$ and  $S_{_Y}$ are determined from an
$\varepsilon_{_\Delta}$-net for $\Delta^g (\alpha)$ as follows.
Let $\{\alpha_{(k)}\}_{1}^q$ be the $\varepsilon_{_\Delta}$-net
for $\Delta^g (\alpha)$, i.e. for any $\alpha\in \Delta^g
(\alpha)$ there is $\alpha_{(k)}$ such that
$$\|\alpha - \alpha_{(k)}\|^2_2\leq \varepsilon_{_\Delta}.
$$
 Let $\x_k =
\x (\cdot,\alpha_{(k)})$ and let $\x$ satisfy the  Lipschitz
condition
$$
\|\x(\cdot,\alpha) - \x (\cdot,\alpha_{(k)})\|^2_{_E}\leq
\lambda_{x} \|\alpha - \alpha_{(k)}\|^2_2
$$
with the Lipschitz constant $\lambda_{x}>0$. Then
$$
\|\x(\cdot,\alpha) - \x (\cdot,\alpha_{(k)})\|^2_{_E}\leq
\lambda_{x}\varepsilon_{_\Delta}.
$$
Let $\varepsilon_{_\Delta}\leq \varepsilon_{_X}/ \lambda_{x}$.
Then $\|\x(\cdot,\alpha) - \x (\cdot,\alpha_{(k)})\|^2_{_E}\leq
\varepsilon_{_X}$ with $\varepsilon_{_X}$ as in Definition 1.
Hence, $S_{_X}=\{\x(\cdot,\alpha_{(k)})\}_{k=1}^p$ is the
$\varepsilon_{_X}$-net for $K_{_X} = \{\x (\cdot,\alpha)
\hspace*{1mm}| \hspace*{1mm}\alpha \in \Delta^g (\alpha)\}$. The
$\varepsilon_{_Y}$-net $S_{_Y}=\{\y(\cdot,\alpha_{(k)})\}_{k=1}^p$
is constructed similarly.

\subsection{Discussion of the solution: associated advantages}

The procedure for constructing the proposed filter (\ref{ff1})
consists of the following steps. First, the operators
$\q_1,\ldots,\q_p$ are chosen so that $\vv_1,\ldots,\vv_p$ are
distinct signals. See Section \ref{choiceqi} in this regard.
Second, the operators $\rrr_1,\ldots,\rrr_p$ are defined as in
Lemma 1. Third, matrices $T_1$, $\ldots,$ $T_p$ are  determined
from Theorem 1. The matrices then satisfy conditions (\ref{int2}).

Such a procedure has the following advantages. First, the
proposed interpolation filter of the $p$th order, $\f_{(p)}$, processes the infinite set of signals
$K_{_Y}$. This advantage has been discussed in Section
\ref{motiv}. Second, the computational effort associated with the
filter $\f_{(p)}$ is facilitated by the orthogonalization procedure
considered in Section \ref{prel}. As a result, each matrix $T_j$
in the filter model $\f_{(p)}$ is determined separately by (\ref{ti})
from equation (\ref{tjew}) and not from a set of matrix equations
as would be required for filters based on a structure similar
to that of (\ref{ff1}) (see, for example, \cite{tor23}). Third,
the filter $\f_{(p)}$ has several free parameters to adjust its
performance. They are the  number of the filter terms, $p$, and
the signal sets $S_{_X}$ and  $S_{_Y}$. See Remarks \ref{rem4} and \ref{rem5} above.

\section{Example of application}

Here, we consider a case when sets $K_{_X}$ and $K_{_Y}$ are finite and large. In the end of Section \ref{int-f}
below, it is shown that this case clearly illustrates its extension to  infinite sets.

{ First, in Section \ref{int-f}, we consider
the interpolation filters presented by (\ref{ff1}) and (\ref{ti}). Then in Section \ref{comp}, we give their comparison with Wiener-type
filters and RLS filters.
}

Let
$K_{_X}=\{\x_1, \x_2,\ldots, \x_N\}$ and $K_{_Y}=\{\y_1, \y_2,\ldots, \y_N\}$, where $N=100$
and $\x_j, \y_j\in L^{2}(\Omega,{\mathbb R}^{n})$  with  $n=116$ for each $j=1,\ldots,8$. Here, $\y_j$ is an observable
random signal (data) and $\x_j$ is a reference random signal (data). That is, $\x_j$ is the signal that should be
estimated by the filter. In this example,
$\x_j$ and  $\y_j$ are simulated as digital images presented by $116 \times 256$ matrices $X_j$
and $Y_j$, respectively. The columns of matrices $X_j$ and $Y_j$ represent a realization of
signals $\x_j$ and $\y_j$, respectively.

Each picture $X_j$ in the sequence $X_1, \ldots, X_N$ has been taken at time $t_j$ with $j=1,\ldots,N$. Time intervals
\begin{equation}\label{dlt}
\Delta_1=t_{2} - t_1, \ldots, \Delta_{N-1}=t_{N} - t_{N-1}
\end{equation}
 were very short. Images $Y_1, \ldots, Y_N$ have been simulated from $X_1, \ldots,
X_N$ in the form $Y_j = X_j\bullet \mbox{\tt rand}_j$ for each $j=1,\ldots,N$. Here, $\bullet$ means
the Hadamard product and $\mbox{\tt rand}_j$ is a $116\times 256$ matrix whose elements are uniformly
distributed in the interval $(0, 1)$.

\begin{center}
\begin{figure}[]
\centering
 \hspace*{-13mm}\begin{tabular}{c@{\hspace*{0mm}}c}
\hspace*{0mm}\vspace*{0mm}\subfigure[Reference data $X_5.$]{\psfig{figure=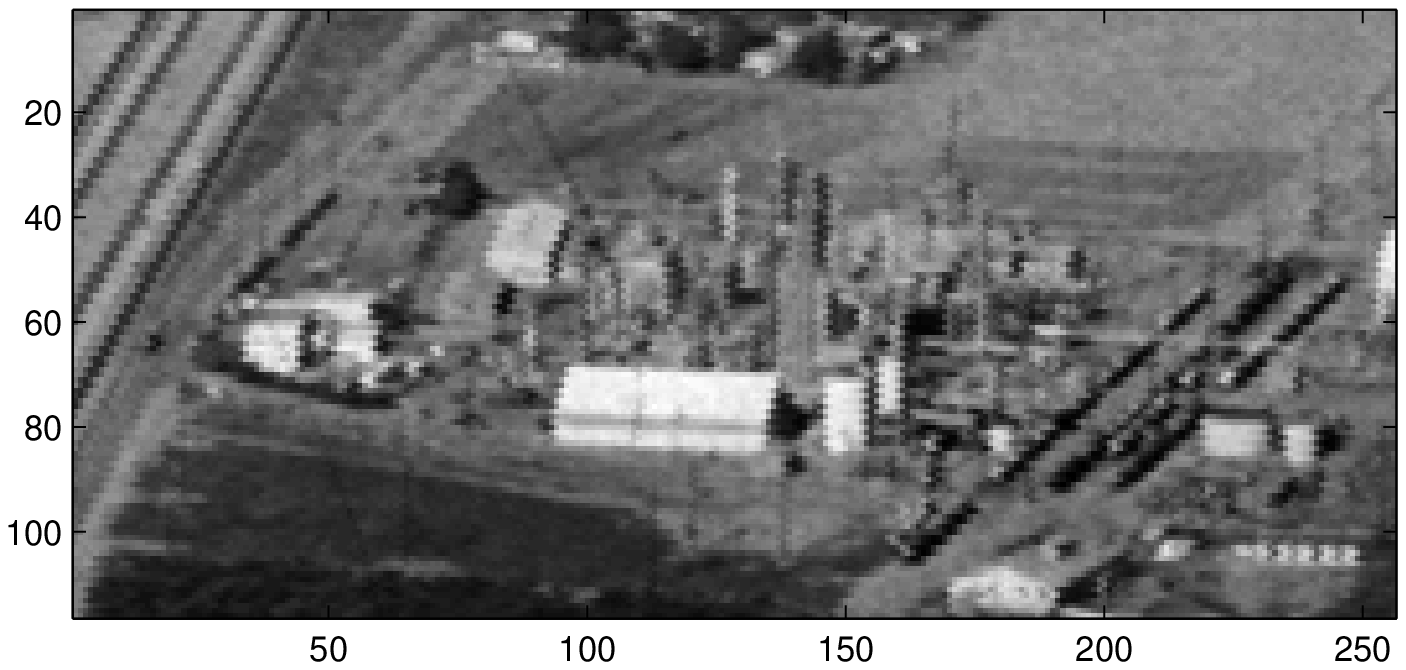,width=8cm,height=4.2cm}}\hspace*{-5mm} &
\hspace*{0mm}\vspace*{0mm}\subfigure[Reference data $X_{18}.$]{\psfig{figure=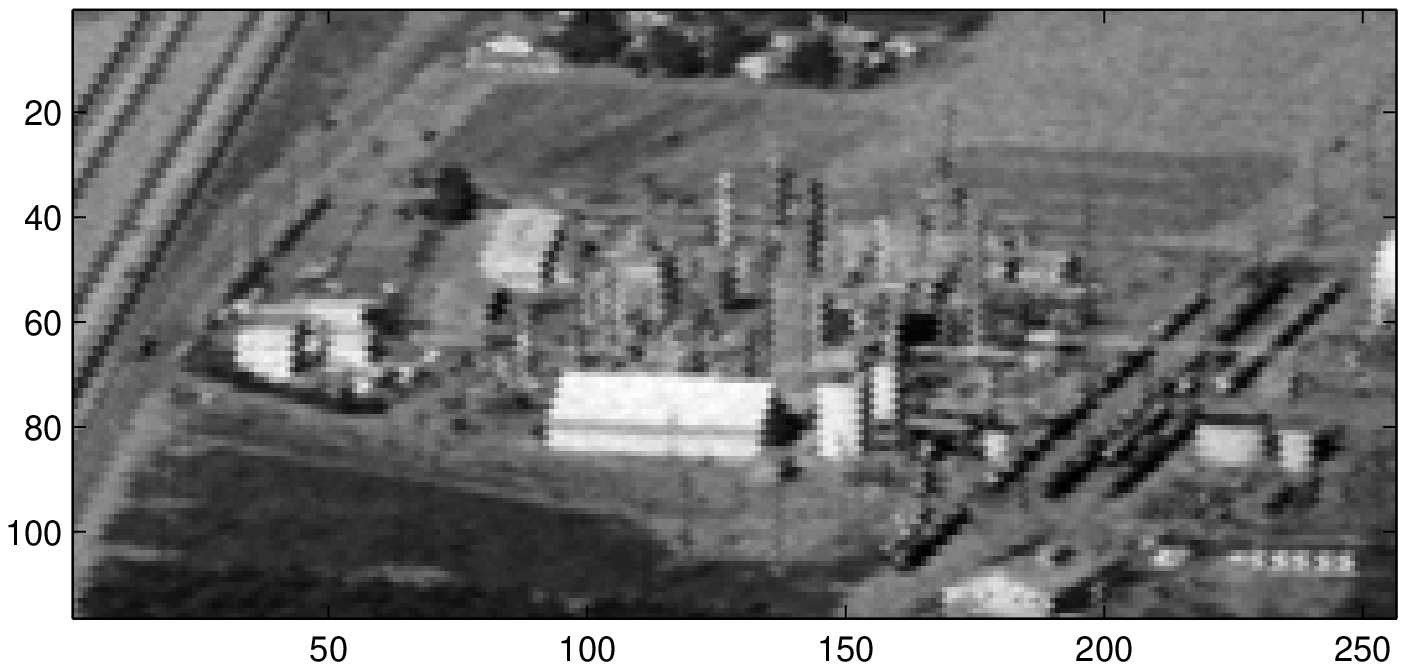,width=8cm,height=4.2cm}}\\
\hspace*{0mm} \vspace*{0mm}\subfigure[Reference data $X_{31}.$]{\psfig{figure=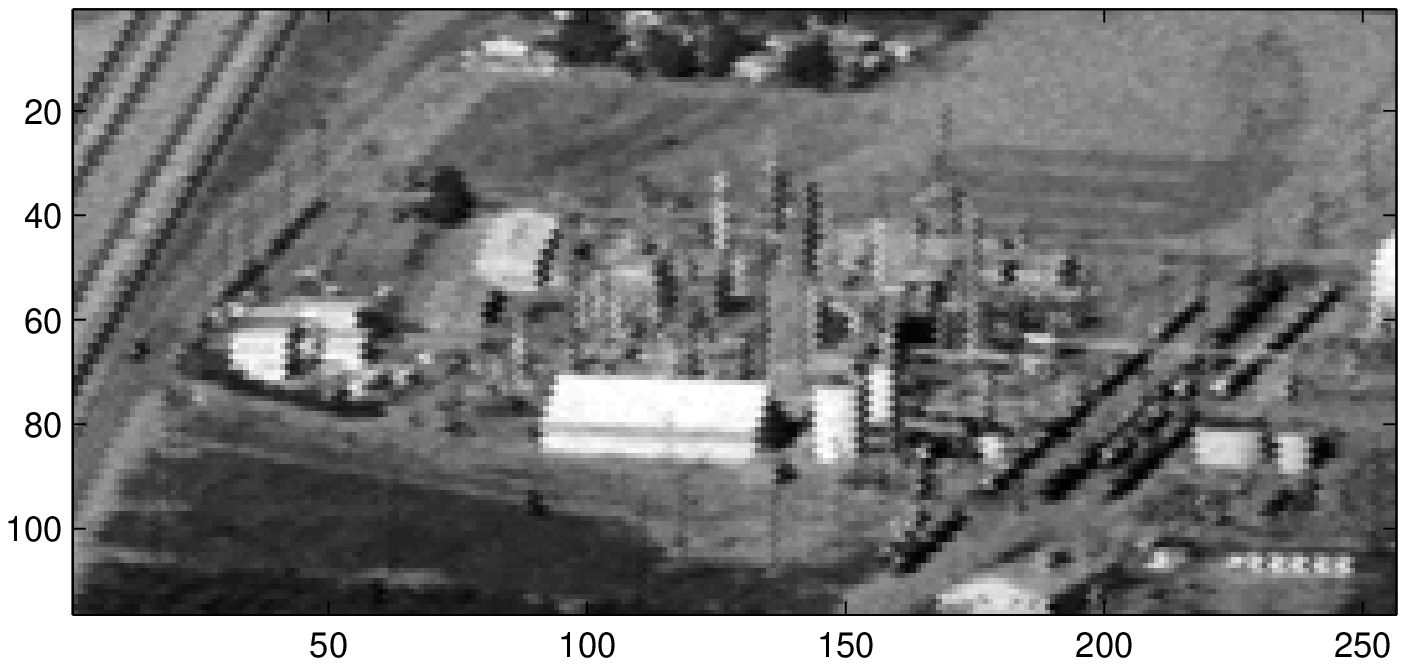,width=8cm,height=4.2cm}}\hspace*{-5mm} &
\hspace*{0mm}\vspace*{0mm}\subfigure[Reference data $X_{44}.$]{\psfig{figure=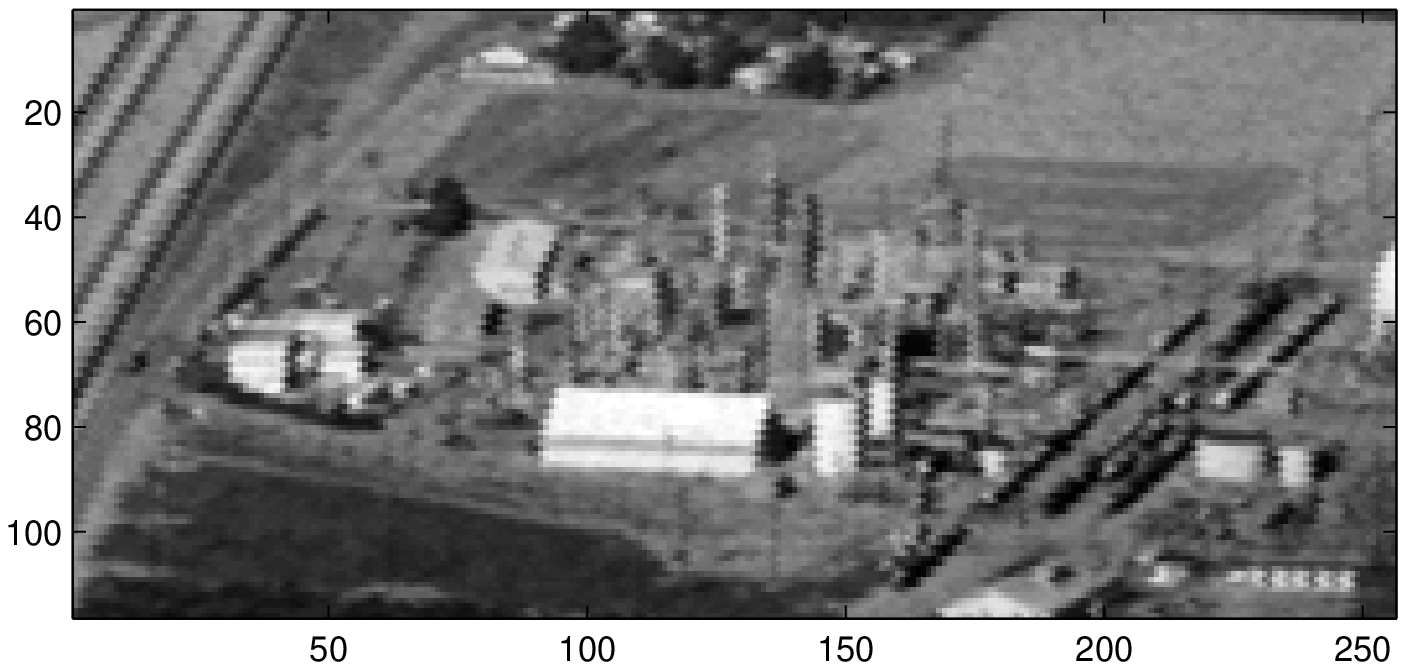,width=8cm,height=4.2cm}}  \\
\hspace*{0mm} \vspace*{0mm}\subfigure[Reference data $X_{57}.$]{\psfig{figure=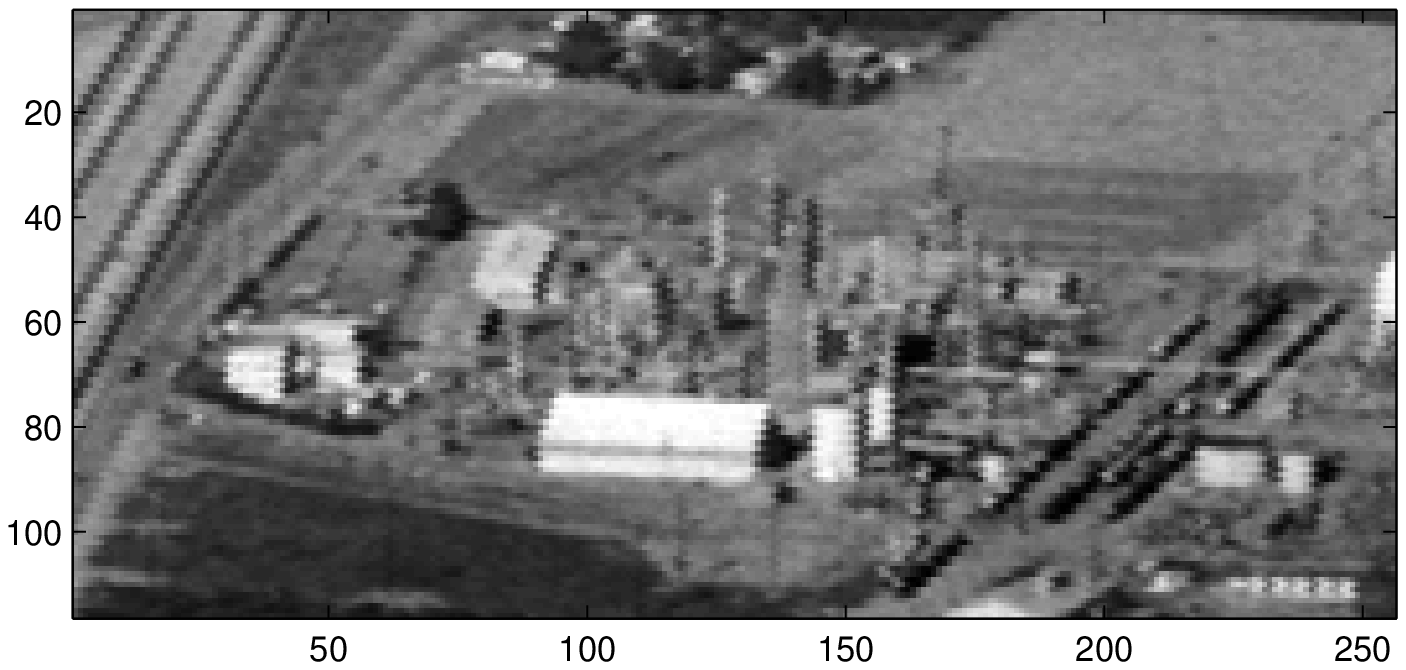,width=8cm,height=4.2cm}}\hspace*{-5mm} &
\hspace*{0mm}\vspace*{0mm}\subfigure[Reference data $X_{70}.$]{\psfig{figure=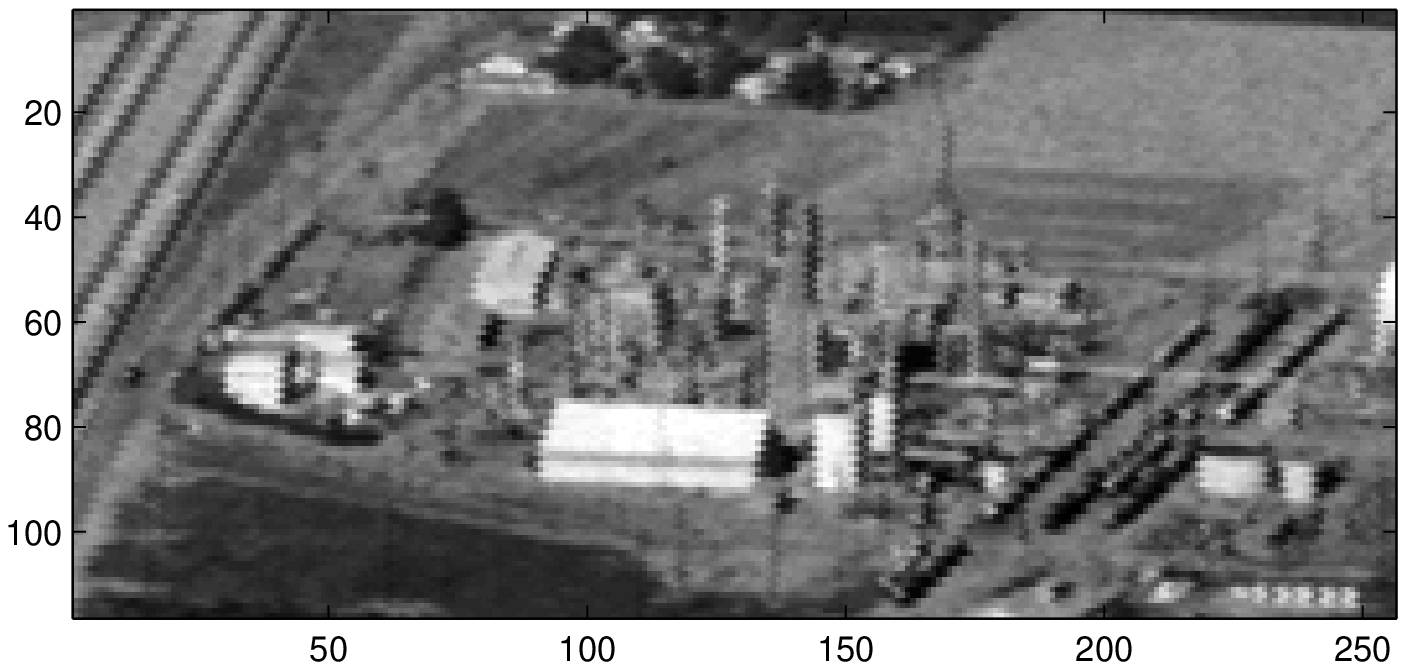,width=8cm,height=4.2cm}}\\
\hspace*{0mm} \vspace*{0mm}\subfigure[Reference data $X_{83}.$]{\psfig{figure=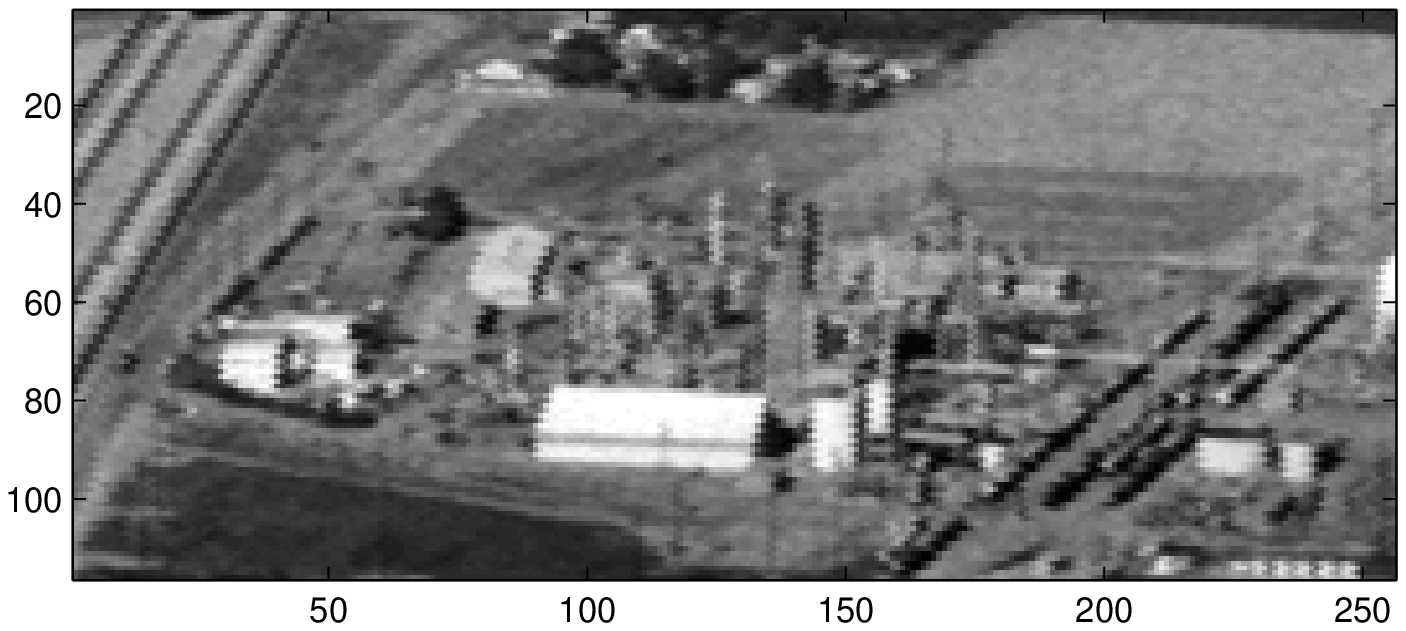,width=8cm,height=4.2cm}}\hspace*{-5mm} &
\hspace*{0mm} \vspace*{0mm}\subfigure[Reference data $X_{96}.$]{\psfig{figure=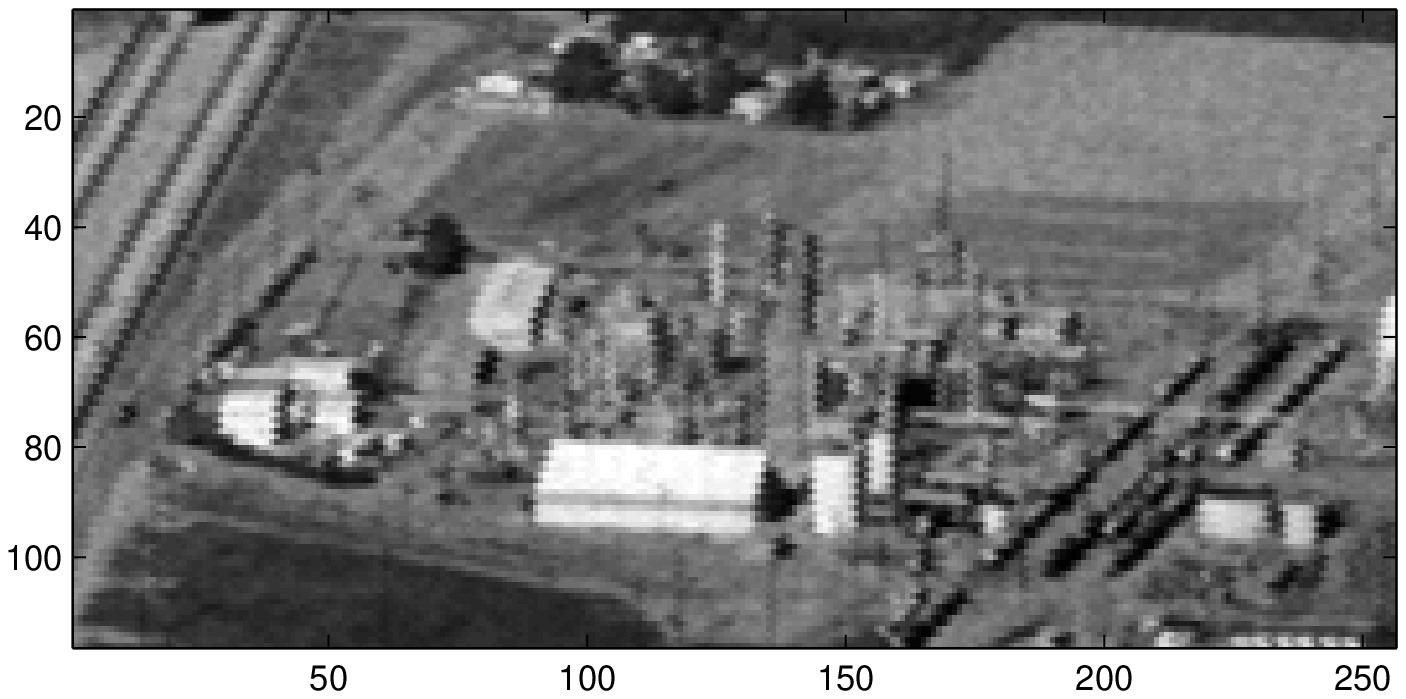,width=8cm,height=4.2cm}}
\end{tabular}
 \vspace*{0mm}\caption{Examples of reference data from set $K_{_X}$   to be estimated from observed data.}
 \label{fig1}
 \end{figure}
\end{center}

To illustrate the sets $K_{_X}$ and $K_{_Y}$ we have chosen their representatives, $X_5$, $X_{18}$, $X_{31}$, $X_{44}$,
$X_{57}$, $X_{70}$, $X_{83}$, $X_{96}$ and $Y_5$, $Y_{18}$, $Y_{31}$, $Y_{44}$, $Y_{57}$, $Y_{70}$, $Y_{83}$, $Y_{96}$,
respectively, taken at times $t_{5 + 13(k-1)}$ with $k=1,\ldots,8$. They are represented in Fig. \ref{fig1} and
Fig. \ref{fig2}.

\subsection{Interpolation filters presented by (\ref{ff1}), (\ref{ti})} \label{int-f}

Sets $S_{_X}$ and $S_{_Y}$ have been chosen in the form
\begin{equation}\label{ssxy}
S_{_X}=\{X_{18}, X_{57}, X_{83}\}\qa S_{_Y}=\{Y_{18},
Y_{57}, Y_{83}\}.
\end{equation}

According to  (\ref{ff1}) and (\ref{ti}), the interpolation filter of the $3rd$ order $F_{(3)}$ that estimates set $K_{_X}=\{X_1, X_2,\ldots, X_{100}\}$ on the basis of observations
$K_{_Y}=\{Y_1, Y_2,\ldots, Y_{100}\}$ is given, for $i=1,\ldots,100$, by
\begin{equation}\label{fyii}
F_{(3)}(Y_i) = T_1R_1Q_1(Y_i) + T_2R_2Q_2(Y_i) +T_3R_3Q_3(Y_i)
\end{equation}
 where $Q_j$, $R_j$ and $T_j$, for $j=1,2,3$, are as follows.

If $i=1,2$, then
\begin{eqnarray}\label{qjyj}
 Q_j(Y_i) = Y_{1-(j-1)}\quad \mbox{ where $j=1,2,3$.}
\end{eqnarray}
If $i=3,\ldots,N$, then
\begin{eqnarray*}
  Q_j(Y_i) = Y_{i-(3-j)} \quad \mbox{ where $j=1,2,3$.}
\end{eqnarray*}
Such a choice is similar to that considered in (\ref{u1}). We denote $V_{ji} =Q_j(Y_i)$ where $j=1,2,3$ and
$i=1,\ldots,N$.

   Matrices $R_j$, with $j=1,2,3$, are determined by Lemma 1 so that, for $i=1,\ldots,100$,
\begin{eqnarray*}
&&\hspace*{-0mm}R_1(V_{1i}) = W_{1i}\quad \mbox{where $W_{1i}:=V_{1i}$},\\
&& \hspace*{-0mm}R_2(V_{2i}) =V_{2i} - K_{21} W_{1i}\quad \mbox{where $ K_{21}=E_{v_{21} w_{1i}}E^\dag_{w_{1i} w_{1i}}$ and
$W_{2i}:=R_2(V_{2i})$}, \\
&& \hspace*{-10mm}\mbox{and}\hspace*{3mm} R_3(V_{3i}) =V_{3i} - K_{231} W_{1i} - K_{32} W_{2i}\quad \mbox{where
$ K_{31}=E_{v_{31} w_{1i}}E^\dag_{w_{1i} w_{1i}}$,}\\
&& \mbox{$ K_{32}=E_{v_{32} w_{2i}}E^\dag_{w_{2i} w_{2i}}$ and}\hspace*{3mm} W_{3i}:=R_3(V_{3i}).
\end{eqnarray*}
Here, we set $M_{21}=\oo$ and $M_{31}=M_{32}=\oo$ (see (\ref{kil})).

 Matrices $T_j$, with $j=1,2,3$, are determined by Theorem \ref{th1}  so that
\begin{eqnarray}\label{tje}
T_j = E_{x_j w_{jj}}E^\dag_{w_{jj} w_{jj}}
\end{eqnarray}
where we have put $A_j=\oo$ (see (\ref{ti})).

Covariance matrices $E_{v_{jk} w_{1\ell i}}$, $E_{w_{\ell i} w_{\ell i}}$ and $E_{x_j w_{jj}}$ used above for certain
$j,k,\ell,i$, can be estimated in various ways. Some of them are presented, for example, in \cite{tor100}, Chapter 4.3.
We note that a covariance matrix estimation is  a specific and difficult problem which is not a subject for this paper.
Here, estimates of $E_{v_{jk} w_{1\ell i}}$, $E_{w_{\ell i} w_{\ell i}}$ and $E_{x_j w_{jj}}$ are used for illustration
purposes only.
Therefore, we use the simplest way to obtain the above estimates, the maximum likelihood estimates \cite{tor100}
based on the  samples $X_i$, $V_{jk}$ and  $W_{\ell i}$. Other related methods, based on incomplete observations, can
be found, e.g., in \cite{per1,tor100}.

\begin{center}
\begin{tabular}{||c|c||c|c||}
\multicolumn{4}{c}{Table 1. Accuracy associated with the proposed interpolation filter of $p$th order.}\\
\hline\hline
\multicolumn{2}{||c||}{ $p=3$ } &  \multicolumn{2}{c||}{ $p=5$ }\\
\hline
 $\|X_{57} - F_{(3)}(Y_{57})\|^2$  &  $\|X_{70} - F_{(3)}(Y_{70})\|^2$   &  $\|X_{57} - F_{(5)}(Y_{57})\|^2$  &  $\|X_{70} - F_{(5)}(Y_{70})\|^2$ \\
\hline
 $9.06\times 10^6$ & $9.12\times 10^6$ & $3.67\times 10^6$ & $3.64\times 10^6$ \\
\hline\hline
\end{tabular}
\end{center}

\bigbreak

\begin{center}
\begin{tabular}{||c|c||c|c||}
\multicolumn{4}{c}{ Table 2. Accuracy associated with  Wiener filters $\it\Phi_j$ and RLS filters $R_j$.}\\
\hline\hline
 $\|X_{57} - {\it\Phi}_{57}(Y_{57})\|^2$ & $\|X_{70} - {\it\Phi}_{70}(Y_{70})\|^2$ & $\|X_{57} - R_{57} (Y_{57})\|^2$ &
 $\|X_{70} - R_{70}(Y_{70})\|^2$   \\
\hline
 $2.34\times 10^8$ & $2.65\times 10^8$ & $9.41\times 10^7$ & $9.77\times 10^7$ \\
\hline\hline
\end{tabular}
\end{center}

\begin{center}
\begin{figure}[]
\centering
 \hspace*{-7mm}\vspace*{0mm}\begin{tabular}{c@{\hspace*{10mm}}c}
\hspace*{0mm}\vspace*{0mm}\subfigure[Observed data $Y_{(57)}.$]{\psfig{figure=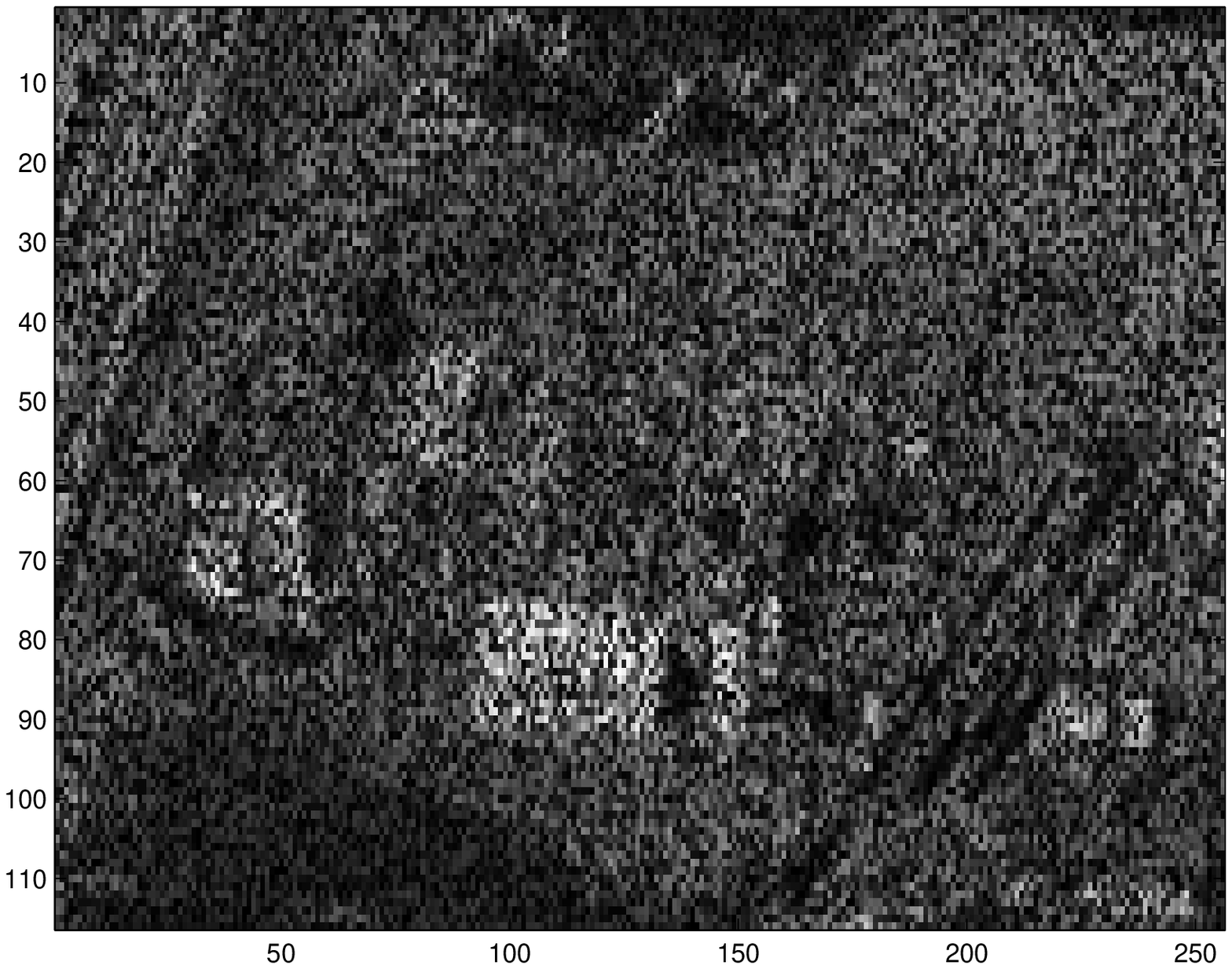,width=6.5cm,height=3.75cm}}\hspace*{0mm} &
\hspace*{0mm}\vspace*{0mm}\subfigure[Observed data $Y_{(70)}.$]{\psfig{figure=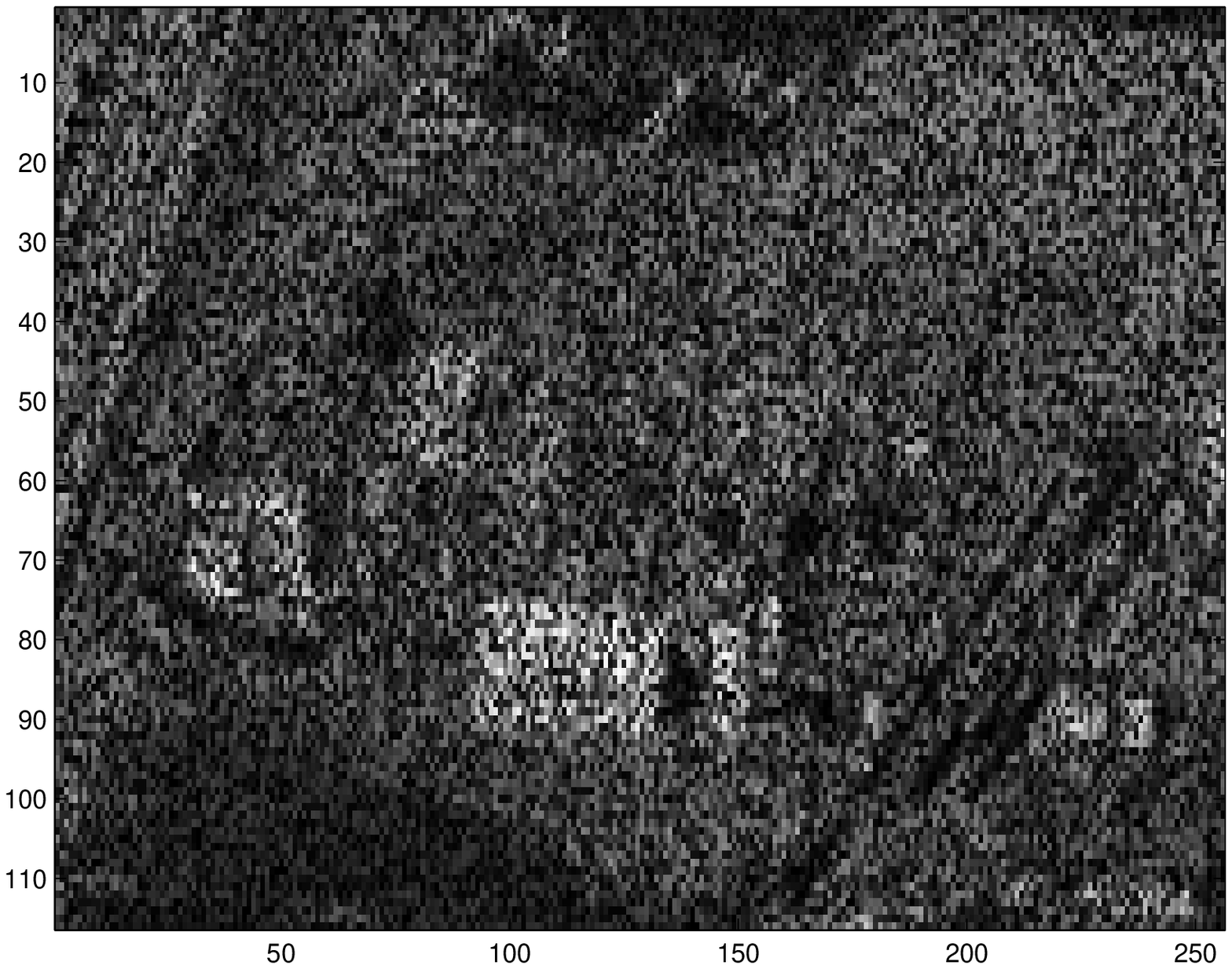,width=6.5cm,height=3.75cm}}\\
\hspace*{0mm} \vspace*{0mm}\subfigure[Estimate $\bar{X}_{57}$ by Wiener filter.]{\psfig{figure=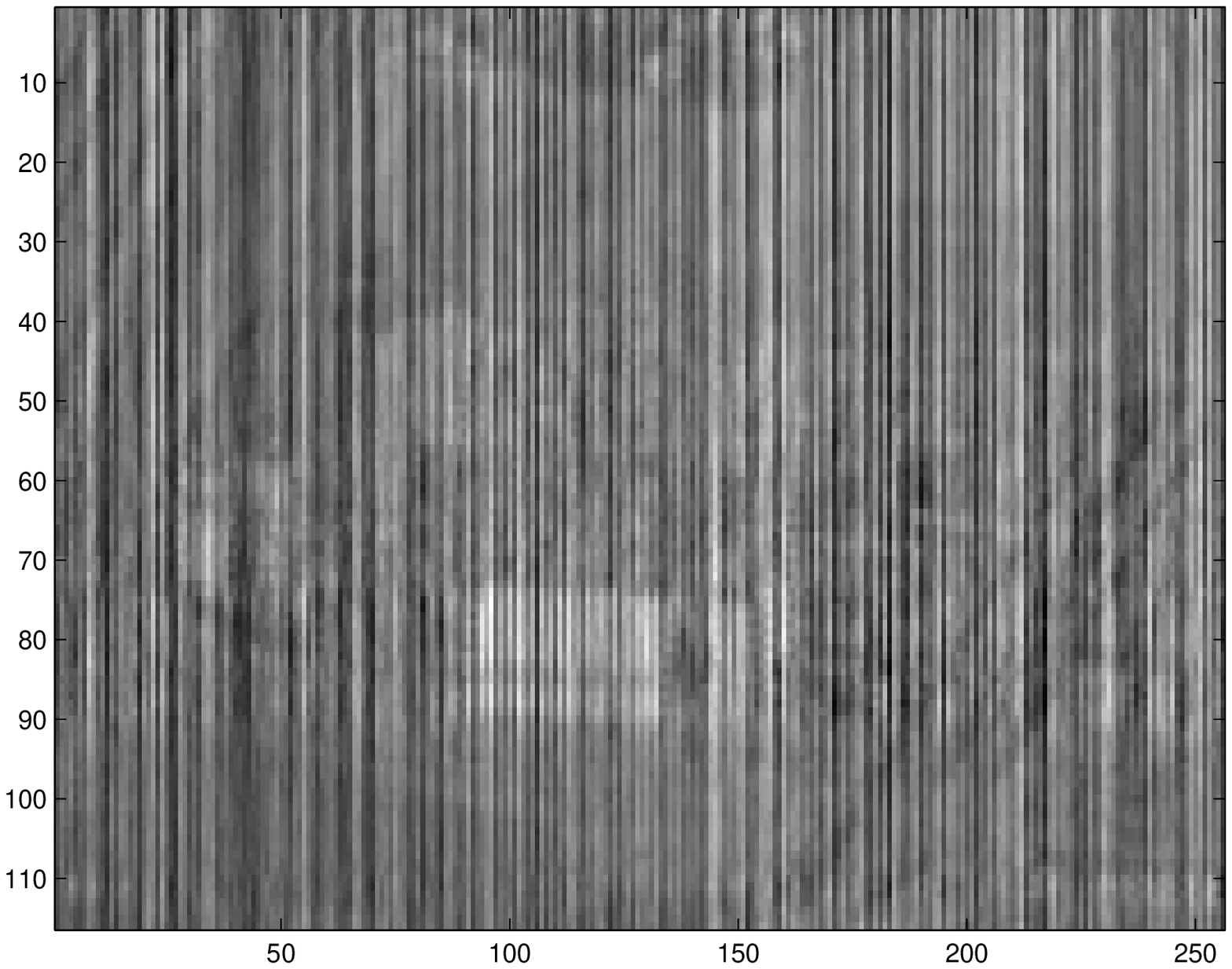,width=6.5cm,height=3.75cm}}\hspace*{0mm} &
\hspace*{0mm}\vspace*{0mm}\subfigure[Estimate $\hat{X}_{70}$ by RLS filter.]{\psfig{figure=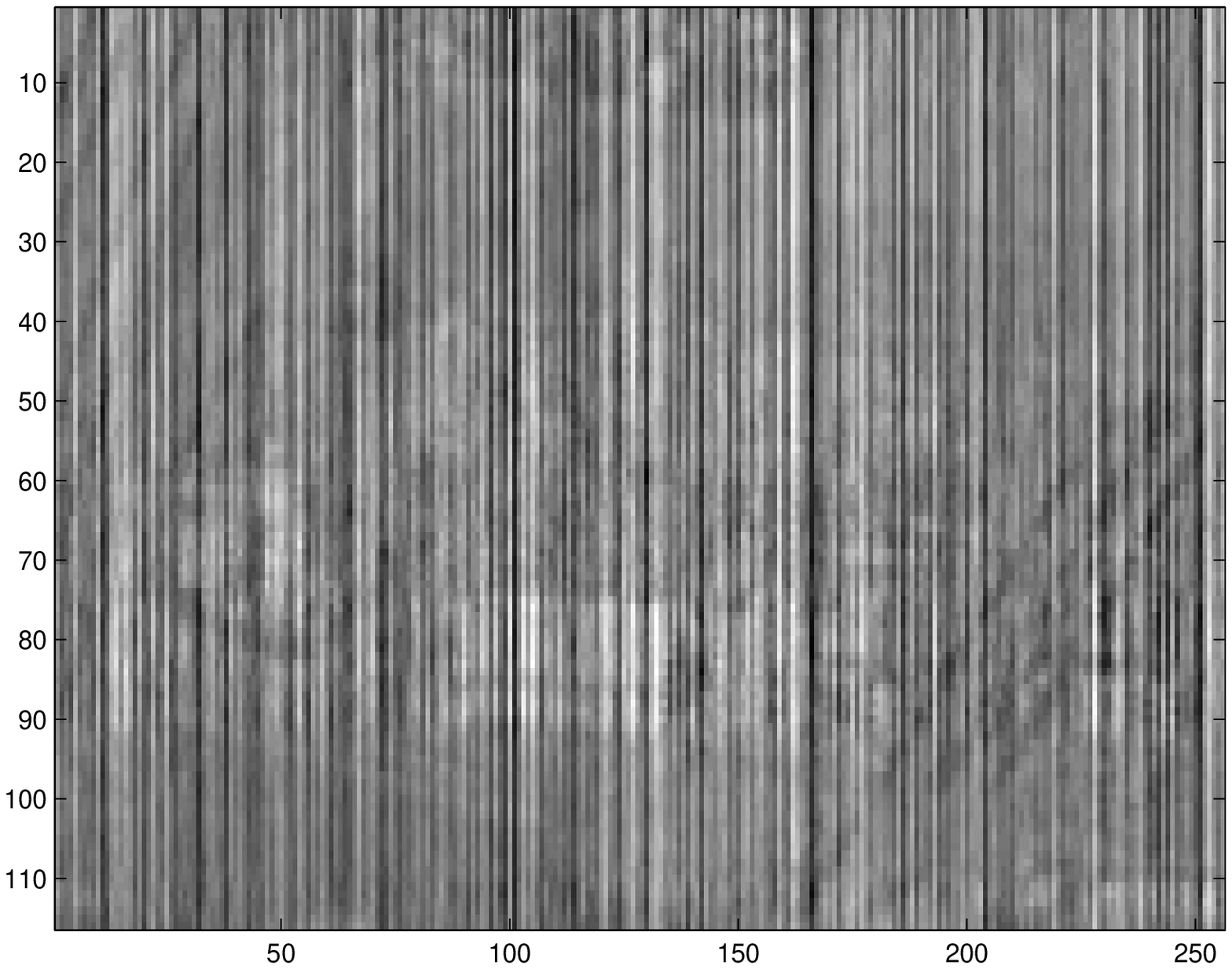,width=6.5cm,height=3.75cm}}  \\
\hspace*{0mm} \vspace*{0mm}\subfigure[Estimate $\widetilde{X}_{57}$ by filter $F_{(3)}$.]{\psfig{figure=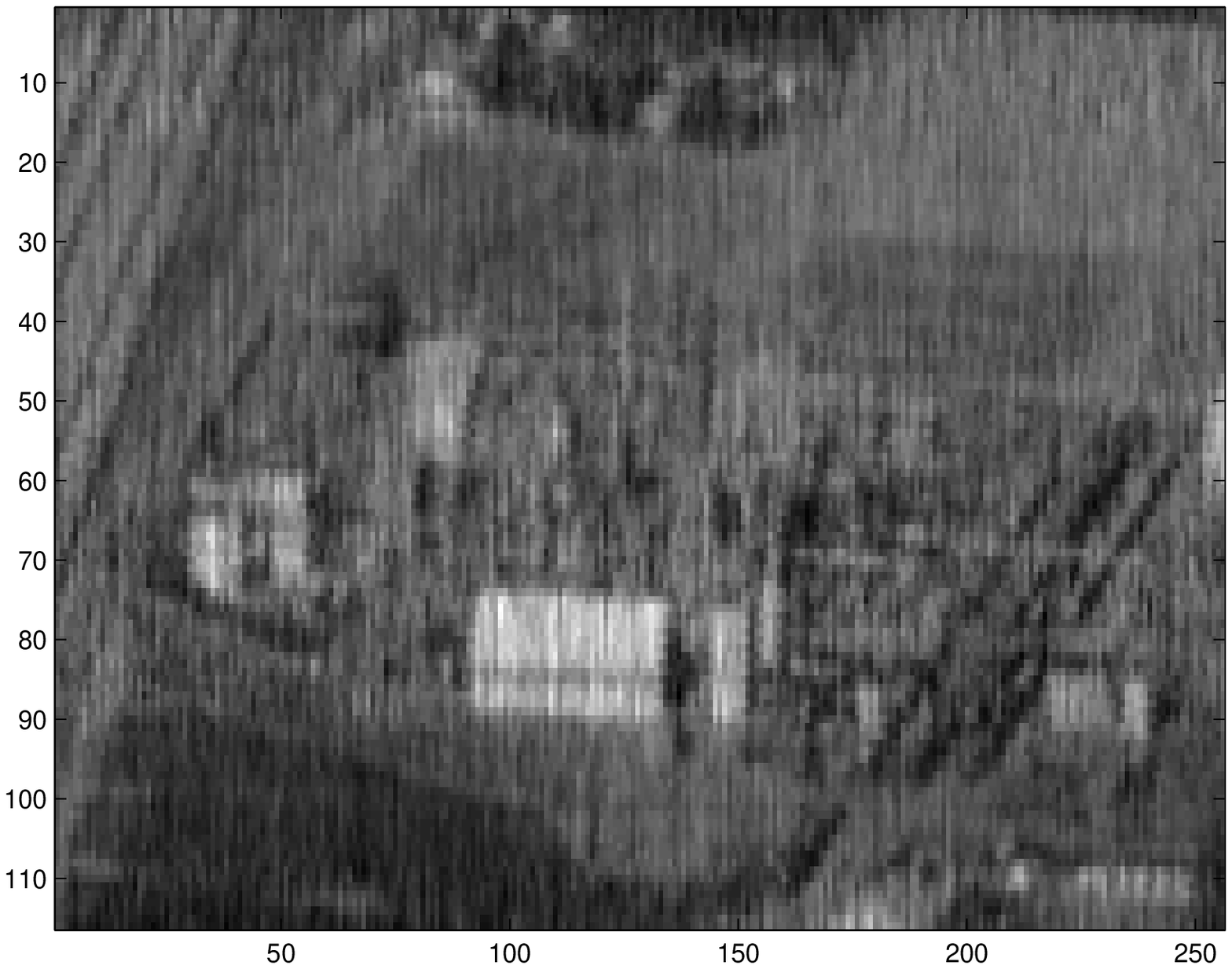,width=6.5cm,height=3.75cm}}\hspace*{0mm} &
\hspace*{0mm}\vspace*{0mm}\subfigure[Estimate $\widetilde{X}_{70}$ by  filter  $F_{(3)}$.]{\psfig{figure=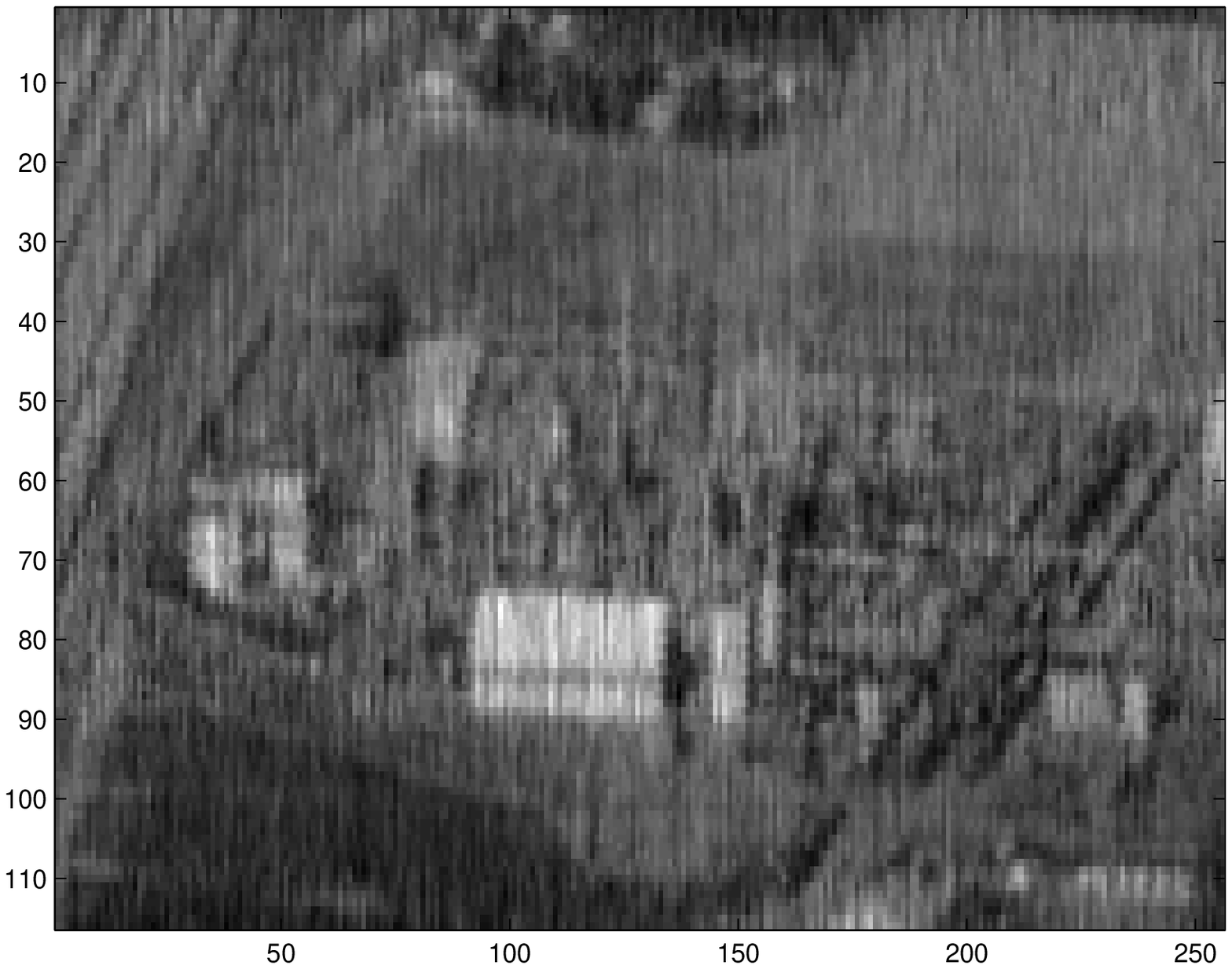,width=6.5cm,height=3.75cm}}  \\
\hspace*{0mm} \vspace*{0mm}\subfigure[Estimate $\widehat{X}_{57}$ by  filter $F_{(5)}$.]{\psfig{figure=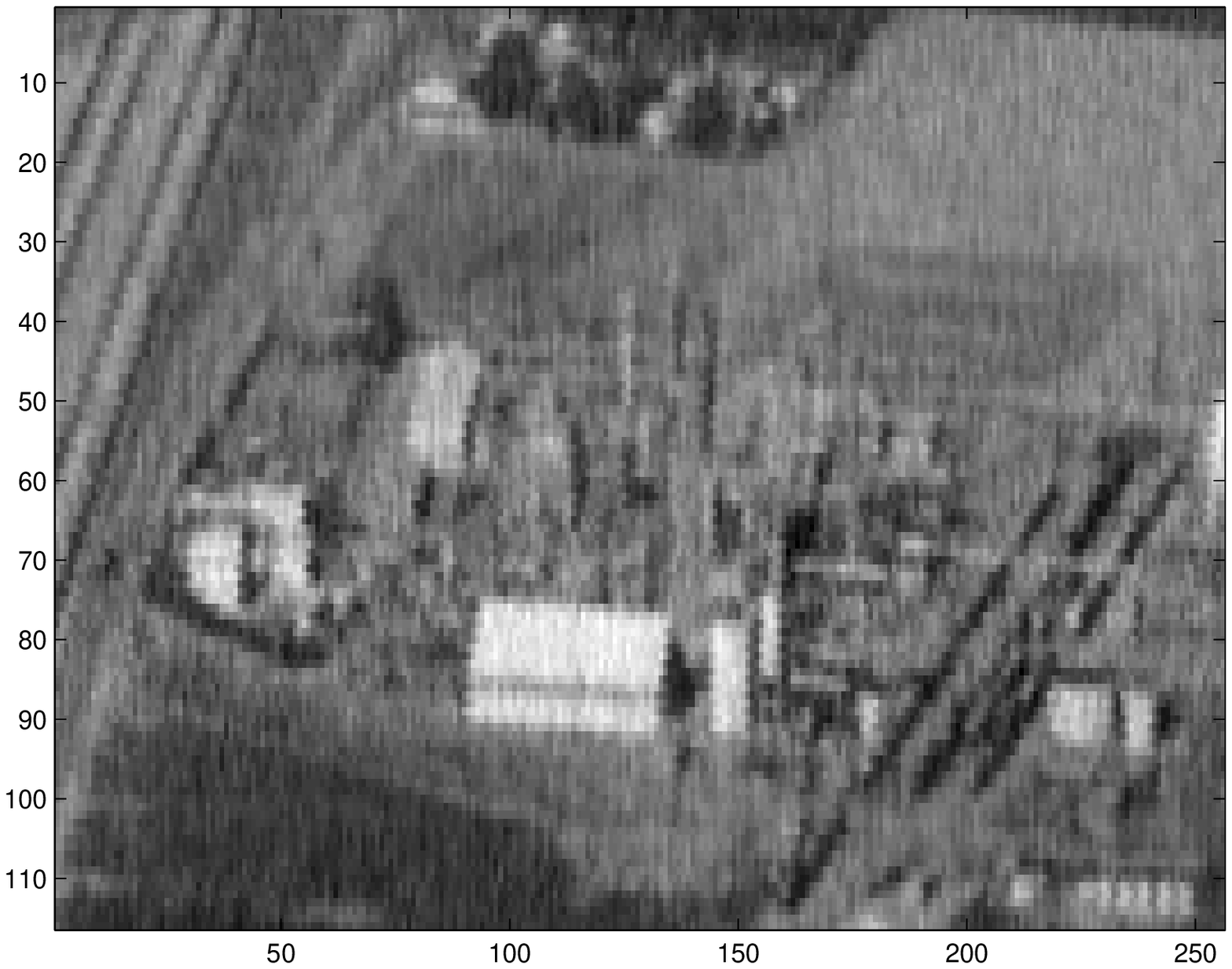,width=6.5cm,height=3.75cm}}\hspace*{0mm} &
\hspace*{0mm}\vspace*{0mm}\subfigure[Estimate $\widehat{X}_{70}$ by  filter $F_{(5)}$.]{\psfig{figure=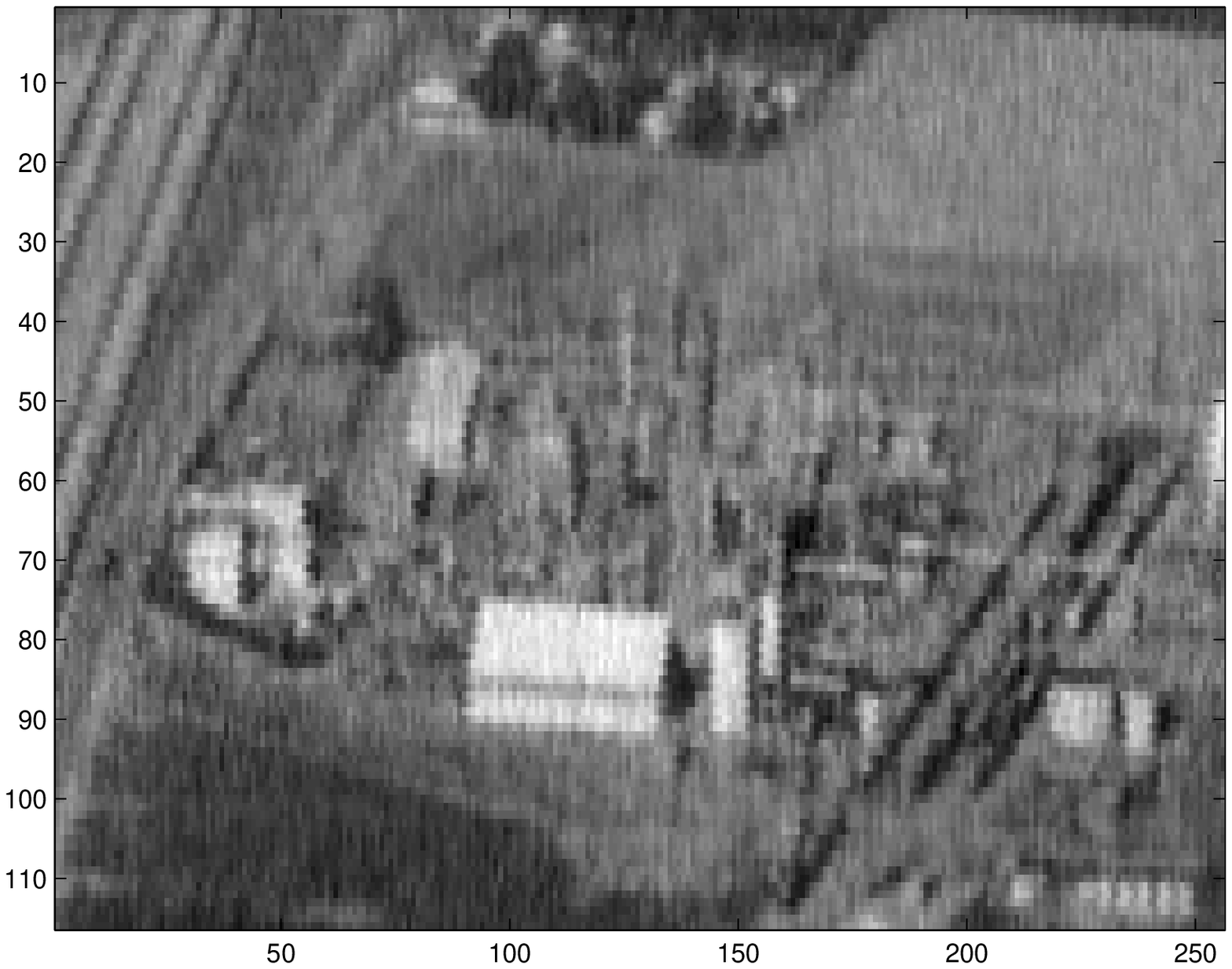,width=6.5cm,height=3.75cm}}
\end{tabular}
 \vspace*{0mm}\caption{Illustration to set $K_{_X}$ estimation by filters $F_{(3)}$ and $F_{(5)}$.}
 \label{fig2}
 \end{figure}
\end{center}

Results of the set $K_{_X}=\{X_1, X_2,\ldots, X_{100}\}$ estimation by the filter given by (\ref{fyii}) are presented in
 Fig. \ref{fig2}.  We represent  typical members of sets $S_{_Y}$ and $K_{_Y}$,
$Y_{57}$ and $Y_{70}$, respectively. Other members of $S_{_Y}$ and $K_{_Y}$ are similar. Estimates of $X_{57}$ and
$X_{70}$ by the filter $F_{(3)}$, $ \widetilde{X}_{57}$ and  $ \widetilde{X}_{70}$,  are also given in Fig. \ref{fig2}.

To illustrate Remark 5, i.e., to show that the error $\|\x - \f_{(p)}(\y)\|^2_{_E}$ decreases if $p$ increases, an estimate of
the set $K_{_X}=\{X_1, X_2,\ldots, X_{100}\}$ has been obtained by the interpolation filter of the $5th$ order $F_{(p)}$, with $p=5$,
i.e. with $p$ bigger than $p=3$
considered in (\ref{fyii}). In this case, $Q_j$, $R_j$ and $T_j$, for $j=1,\dots,5$ are determined in  ways  similar to
(\ref{qjyj})--(\ref{tje}). We denote  $ \widehat{X}_{i}=F_{(5)}Y_{i}$, for $i=1,\ldots,N$. Estimates $\widehat{X}_{57}$ and $\widehat{X}_{70}$
are given in Fig. \ref{fig2}.

In Table 1, values of the errors $\|X_i - F_{(p)}Y_i\|^2$ are presented for $i=57$ and $i=70$, and $p=3$ and $p=5$.

It follows from Table 1 and Fig.  \ref{fig2} that an increase in $p$ implies a decrease in the error of the estimation.

Note that, although the considered sets $K_{_X}$ and $K_{_Y}$ are finite, the above results of their estimation by the interpolation
filters  can easily be extended to the case when $K_{_X}$ and $K_{_Y}$ are infinite.  Indeed, if in (\ref{dlt}), $\Delta_i \rightarrow 0$ as
$i \rightarrow \infty$, then $K_{_X}$ and $K_{_Y}$ tend to infinite sets. In this case, the choice of sets $S_{_X}$ and $S_{_Y}$ can be the same
as in (\ref{ssxy}) above. Then filters $F_{(3)}$ and $F_{(5)}$ will also be the same as above. Therefore, for the case when
$K_{_X}$ and $K_{_Y}$ are infinite sets,  estimates of $K_{_X}$ by filters $F_{(3)}$ and $F_{(5)}$ are similar to the results obtained
above.

\subsection{Comparison with Wiener-type filter and RLS filter} \label{comp}

{ In both the  Wiener filtering approach and the RLS filtering methodology, a  filter should be found for each pair of input-output
signals,
$\{\x_j, \y_j\}$  where $j=1,\ldots,100$. That is, in these simulations,  $100$ Wiener filters or RLS filters, $\it\Phi_{1},\ldots,\it\Phi_{100}$
or  $R_{1},\ldots,R_{100}$, respectively,  are required to process signals from $K_{_Y}$ to $K_{_X}$.
In contrast, our approach requires the single interpolation filter $\f_{(p)}$ with $p$ terms (above, we have chosen $p=3$ and $p=5$) where each term
requires computational effort that
 is similar to that required by the single Wiener filter $\it\Phi_j$ and the single RLS filter $R_j$. Clearly, the Wiener filtering approach and the RLS filtering
methodology imply much  more computational work \cite{hay1,say1} to process  signals from $K_{_Y}$ to $K_{_X}$.

Examples of accuracies associated with the  filters $\it\Phi_{57}$ and $R_{70}$ for estimating $X_{57}$ and $X_{70}$ (which represent
 typical signals under consideration)  are given in Table 2, where $\bar{X}_{57}=\it\Phi_{57}(Y_{57})$ and $\hat{X}_{70}
 =R_{70}(Y_{70})$. The estimates $\bar{X}_{57}$ and $\hat{X}_{70}$ are presented in  Fig. \ref{fig2}.

The accuracy associated with the proposed interpolation filter $\f_{(p)}$ is better than that of filters $\it\Phi_j$ and $R_j$  due to
the  properties of
filter $\f$  discussed above (in particular, in Theorems \ref{th2} and \ref{th3}).

}

\section{Conclusion}

In this paper, we develop a new approach to filtering of infinite
sets of stochastic signals. The approach is based on a filter
representation in the form of a sum of $p$ terms. Each term is
derived from three matrices, $Q_i$, $R_i$ and $T_i$ with
$i=1,\ldots,p$. The methods for determining these matrices have
been studied. In particular, matrices $T_i,\ldots,T_p$ are
determined from interpolation conditions. This procedure allows us
to construct the filter that estimates each signal from the
given infinite set with  controlled accuracy.

An analysis of the error associated with the filter has been
provided. The analysis has shown that the filter has three
free parameters with which to improve performance. It follows
from the error analysis that the proposed filter is
asymptotically optimal. The filter is determined in terms of
pseudo-inverse matrices and, therefore, it always exists.

\end{document}